\global\def\draftcontrol{0}  
  
   \def\versionno{D3/D7 on the conifold}  
  
\catcode`\@=11  
  
\expandafter\ifx\csname draftcontrol\endcsname\relax\global\def\draftcontrol{0}  
\fi  
  
{\count255=\time\divide\count255 by 60  
\xdef\hourmin{\number\count255}  
\multiply\count255 by-60\advance\count255 by\time  
\xdef\hourmin{\hourmin:\ifnum\count255<10 0\fi\the\count255}}  
\def\draftdate{\number\month/\number\day/\number\year\ \ \ \hourmin }

\newcommand\makepapertitle{\par  
  \begingroup  
    \renewcommand\thefootnote{\@fnsymbol\c@footnote}%
    \def\@makefnmark{\rlap{\@textsuperscript{\normalfont\@thefnmark}}}%
    \long\def\@makefntext##1{\parindent 1em\noindent  
            \hb@xt@1.8em{%
                \hss\@textsuperscript{\normalfont\@thefnmark}}##1}%
     \newpage  
     \global\@topnum\z@   
     \@makepapertitle  
     \thispagestyle{empty}\@thanks  
  \endgroup  
  \setcounter{footnote}{0}%
  \global\let\thanks\relax  
  \global\let\makepapertitle\relax  
  \global\let\@makepapertitle\relax  
  \global\let\@thanks\@empty  
  \global\let\@author\@empty  
  \global\let\@date\@empty  
  \global\let\@title\@empty  
  \global\let\title\relax  
  \global\let\author\relax  
  \global\let\date\relax  
  \global\let\and\relax  
  \def\version{\let\version\@version\@gobble}  
}  
\def\@makepapertitle{%
  \newpage  
   \ifnum\draftcontrol=1 {}  
   \version\versionno  
   \vskip 3em%
   \else  
   \hfill\hbox to 3cm {\parbox{4cm}{\@pubnum}\hss}%
   \vskip 3em%
   \fi  
   \begin{center}%
   \let \footnote \thanks  
     {\LARGE {\@title}}%
     \vskip 1.5em%
     {\normalsize
       \lineskip .5em%
       \begin{tabular}[t]{c}%
         \@author  
       \end{tabular}\par}%
     \vskip 1.5em%
     {\@bstract}%
     \end{center}%
     \vskip 1.5em  
     \@date%
   \par  
}  
  
\gdef\@pubnum{}  
\def\pubnum#1{%
  \gdef\@pubnum{#1}}  
  
\gdef\@bstract{}  
\def\Abstract#1{%
  \gdef\@bstract{%
   \parbox{\textwidth-0pc}{%
   \centerline{\bf Abstract}\penalty1000%
\noindent
\renewcommand\baselinestretch{1.0}%
{#1}}}  
}

\def\ps@paper{\let\@mkboth\@gobbletwo%
     \ifnum\draftcontrol=1  
        \def\@oddfoot{\hbox to \textwidth{\tiny \versionno \hfil\tiny\draftdate}%
        \hskip -\textwidth \hbox to \textwidth{\hfil\rm\thepage\hfil}}%
     \else\def\@oddfoot{\hbox to \textwidth{\hfil\rm\thepage\hfil}}  
     \fi  
     \let\@evenfoot\@oddfoot  
}  
  
\def\@version#1{\ifnum\draftcontrol=1  
\typeout{}\typeout{#1}\typeout{}  
\vskip3mm\centerline{\hbox{\fbox{\normalsize{\tt DRAFT -- #1 -- }  
                   {\draftdate}}}}\vskip3mm  
\fi}  
\let\version\@version  
\long\def\eqlabel#1{\ifnum\draftcontrol=1  
                    \tag@false  
                    \tag*{(\theequation) \hbox to -0.2cm{\hspace{0cm}\small{#1}\hss}}  
                    \refstepcounter{equation}  
                    \edef\@currentlabel{\theequation}  
                    \ltx@label{#1}          
                    \else  
                    \label{#1}  
                    \fi  
                    }  
\let\st@bibitem\@bibitem  
\let\st@lbibitem\@lbibitem  
\ifnum\draftcontrol=1  
  \def\@bibitem#1{%
    \st@bibitem{#1}\a@@label{#1}\ignorespaces}  
  \def\@lbibitem[#1]#2{%
    \st@lbibitem[#1]{#2}\a@@label{#2}\ignorespaces}  
  \def\a@@label#1{%
    \gdef\a@lab{\smash{\normalfont\small#1}}  
    \ifvmode  
      \if@inlabel  
        \global\setbox\@labels\hbox{%
          \llap{\a@lab\let\a@lab\relax  
                \kern\@totalleftmargin\kern\marginparsep}%
          \box\@labels}%
      \fi  
    \fi}  
\fi  

\documentclass[12pt,letterpaper]{article}  
  
\usepackage{amsmath,amssymb,array,calc,rotating,epsfig,psfrag}  
\usepackage[nosort]{cite}  
  
\ifnum\draftcontrol=1  
\fi  
  
\renewcommand\baselinestretch{1.25}  
\renewcommand\section{\@startsection {section}{1}{\z@}%
                                   {-3.5ex \@plus -1ex \@minus -.2ex}%
                                   {2.3ex \@plus.2ex}%
                                   {\normalfont\large\bfseries}}  
\renewcommand\subsection{\@startsection{subsection}{2}{\z@}%
                                   {-3.25ex\@plus -1ex \@minus -.2ex}%
                                   {1.5ex \@plus .2ex}%
                                   {\normalfont\normalsize\bfseries}}  
\renewcommand\subsubsection{\@startsection{subsubsection}{3}{\z@}%
                                   {-3.25ex\@plus -1ex \@minus -.2ex}%
                                   {1.5ex \@plus .2ex}%
                                   {\normalfont\normalsize\it}}  
\renewcommand\paragraph{\@startsection{paragraph}{4}{\z@}%
                                   {-3.25ex\@plus -1ex \@minus -.2ex}%
                                   {1.5ex \@plus .2ex}%
                                   {\normalfont\normalsize\bf}}  
  
\def\revise#1       {\raisebox{-0em}{\rule{3pt}{1em}}%
                     \marginpar{\raisebox{.5em}{\vrule width3pt\  
                     \vrule width0pt height 0pt depth0.5em  
                     \hbox to 0cm{\hspace{0cm}{%
                     \parbox[t]{4em}{\raggedright\footnotesize{#1}}}\hss}}}}

\def\calf         {{\cal F}}

\def\caln         {{\cal N}}

\def\calw         {{\cal W}}

\def\reals        {{\mathbb R}}

\def\del          {\partial}

\def\Im           {{\rm Im\hskip0.1em}}

\def\sqr#1#2{{\vcenter{\vbox{\hrule height.#2pt  
 \hbox{\vrule width.#2pt height#1pt \kern#1pt  
 \vrule width.#2pt}\hrule height.#2pt}}}}

\newcommand{\fft}[2]{{\frac{#1}{#2}}}  
\newcommand{\ft}[2]{{\textstyle{\frac{#1}{#2}}}}

\def\a{\alpha}  
  
\def\r{\rho}

\def\m{\mu}  
\def\g{\gamma}  
\def\l{\lambda}  
\def\n{\nu}  
\def\bn{\bar{\nu}}

\catcode`\@=12

\newcommand{\be}{\begin{equation}}  
\newcommand{\ee}{\end{equation}}  
\newcommand{\beq}{\begin{equation}}  
\newcommand{\eeq}{\end{equation}}  
\newcommand{\ba}{\begin{eqnarray}}  
\newcommand{\ea}{\end{eqnarray}}  
  
\def\del{{\partial}}

\def\te{\theta}

\def\lbldef#1#2{\expandafter\gdef\csname #1\endcsname {#2}}

\def\href#1#2{#2}

\newcommand{\ber}{\begin{eqnarray}}  
\newcommand{\eer}{\end{eqnarray}}  
  
\newcommand{\bea}{\begin{eqnarray}}  
\newcommand{\eea}{\end{eqnarray}}

\newcommand{\beqar}{\begin{eqnarray}}

\newcommand{\eeqar}{\end{eqnarray}}  

\newcommand{\dsl}  
  {\kern.06em\hbox{\raise.15ex\hbox{$/$}\kern-.56em\hbox{$\partial$}}}  

\newcommand{\D}{{\cal{D}}}

\newcommand{\eeqarr}{\end{eqnarray}}  
\newcommand{\ZZ}{{\rm \kern 0.275em Z \kern -0.92em Z}\;}

\def\a{\alpha}  
  
\def\r{\rho}  
\def\d{\delta}  
\def\g{\gamma}

\makeatletter \@addtoreset{equation}{section} \makeatother  
\renewcommand{\theequation}{\thesection.\arabic{equation}}  

\def\be{\begin{equation}}  
\def\ee{\end{equation}}  
\def\bea{\begin{eqnarray}}  
\def\eea{\end{eqnarray}}  
\def\m{\mu}  
\def\n{\nu}  
\def\g{\gamma}  
\def\p{\phi}  
\def\L{\Lambda}  
\def \W{{\cal W}}  
\def\bn{\bar{\nu}}  
  
\def\bw{\bar{w}}  
\def\ba{\bar{\alpha}}  
  
\def\bz{\bar{z}}

\begin{document}  
\begin{titlepage}  
  
\version\versionno  
  
\hbox to \hsize{{\tt hep-th/0406207}\hss  
\vtop{\hbox{MCTP-04-34}  
\hbox{PUPT-2124}  
}}

\vskip 1.7 cm

\centerline{\bf \Large Holographic Duals of Flavored $\caln=1$ Super  
Yang-Mills:}  
\vskip .4cm  
\centerline{\bf \Large Beyond the Probe Approximation}  
  
\vskip .5cm  
  
\centerline{Benjamin A. Burrington$^1$, James T. Liu$^1$,  
Leopoldo A. Pando Zayas$^1$ and Diana Vaman$^2$}  
  
\vskip .2cm  
\centerline{\it ${}^1$ Michigan Center for Theoretical Physics}  
\centerline{ \it Randall Laboratory of Physics, The University of Michigan}  
\centerline{\it Ann Arbor, MI 48109-1120}  
\vskip .2cm  
\centerline{\it ${}^2$ Department of Physics, Princeton University}  
\centerline{\it Princeton, NJ 08544}  
  
\vspace{.4cm}  
  
\begin{abstract}  
We construct backreacted D3/D7 supergravity backgrounds which are dual to  
four-dimensional $\caln=1$ and $\caln=2$ supersymmetric Yang-Mills at large  
$N_c$ with flavor quarks in the fundamental representation of  
SU$(N_c)$.  We take into account the backreaction of D7-branes on either  
AdS$_5 \times S^5$ or AdS$_5 \times T^{1,1}$, or more generically on  
backgrounds where the space transverse to the D3-branes is K\"ahler.  
The construction of the backreacted geometry splits into two stages.  
First we determine the modification of the six-dimensional space  
transverse to the D3 due to the D7, and then we compute the warp factor  
due to the D3.  
  
The $\caln=2$ background corresponds to placing a single stack of $N_f$  
D7-branes in AdS$_5 \times S^5$. Here the K\"ahler potential is known  
exactly, while the warp factor is obtained in certain limits as a  
perturbative expansion.  By placing another D7$'$ probe in the backreacted  
D3/D7 background, we derive the effect of the D7-branes on the spectrum of  
the scalar fluctuations to first order in $N_f$.  
The two systems with $\caln=1$ supersymmetry that we discuss are  
D3/D7/D7$'$ and D3/D7 on the conifold. In both cases, the K\"ahler  
potential is obtained perturbatively in the number of D7-branes. We  
provide all the ingredients necessary for the computation of each term  
in the expansion, and in each case give the first few terms explicitly.  
Finally, we comment on some aspects of the dual gauge theories.  
  
\end{abstract}

  

\end{titlepage}  
  
  
  
\section{Introduction}  
In the framework of the gauge/gravity correspondence \cite{magoo} a complete  
holographic description of QCD is still beyond our current  
reach. Nevertheless, substantial progress has been achieved in the  
holographic understanding of pure $\caln=1$ super Yang-Mills (SYM) at  
large $N_c$ \cite{mn,ks}.  
The holographic understanding of flavor dynamics at the quantitative  
level has developed slower, in  
great measure due to the lack of a fully backreacted supergravity  
background describing gauge theories with fundamental flavor. In this  
paper we address the construction of fully backreacted  
supergravity backgrounds dual to $\caln=1$ SYM with fundamental flavor.  
  
Recently, there has been much attention devoted to flavor dynamics in  
the probe approximation limit  
\cite{kk,martin,horatiu,peter,carlos,martin1,martin2,cobi,evans,evans2,evans3,italia,schnitzer}.  
However, a significant limitation of the  
probe approximation is that in that regime the number of flavors is necessarily  
much smaller than the number of colors: $N_f/N_c \ll 1$. Still,  
even in the probe approximation, there have been some quite interesting  
results \cite{etaprime1,etaprime2}.  
  
{\it Why go beyond the probe approximation?}  
One of the major stepping stones in a holographic understanding of QCD  
is a control of  flavor dynamics. For example,  
dynamical mesons and baryons play a central r\^ole in obtaining  
a complete understanding of the QCD phase diagram.  
In particular, in  
the large $N_c$ limit the critical temperature should behave very  
differently in the presence of baryons. On the other hand, it is known  
that baryons play a subleading r\^ole at large $N_c$, which  
leads to the question of how precisely they change the critical  
temperature. On a more formal level, the quantum moduli space of  
SU$(N_c)$ supersymmetric gauge theories with $N_f$ flavors crucially  
depends on the relative  value of $N_c$ and $N_f$.  There is a host of  
phenomena in  flavor dynamics of supersymmetric theories, such as the  
conformal window,  that can be  
explored only in the limit of comparable $N_c$ and $N_f$. Having $N_f  
\ll N_c$ excludes, in principle, the possibility of a holographic  
description of such phenomena. Clearly,  
unless a fully backreacted solution becomes available, a quantitative  
understanding of several physical questions of gauge theories will remain  
elusive in the holographic approach.  
  
In this paper we address the question of backreaction  in the  
context of supergravity backgrounds dual to $\caln=1$ supersymmetric  
gauge theories.  
While the general class of supergravity backgrounds dual to $\caln=1$ gauge  
theories is wide, we concentrate only on backgrounds  
containing D3 and D7-branes. In this case the backreacted metric of  
the space transverse to the stack of D3-branes remains K\"ahler.  
For K\"ahler  
metrics a remarkable simplification of the  Einstein equations takes  
place. In fact, solving the Einstein equations is equivalent to solving an  
equation  that in many cases becomes a standard Monge-Amp\`ere  
equation. Therefore, the problem of finding backreacted solutions  
splits into two main steps. In the first one, after finding a suitable  
Ansatz for the dilaton, we determine the K\"ahler potential of the resulting  
six-dimensional background transverse to the D3 worldvolume. In the second  
step we find the warp factor due to the presence of the D3-branes.  
  
The paper is organized as follows. In section~\ref{d3d7} we review the  
construction of the D3/D7 system to introduce notation and comment on  
some of the choices that we make further in the paper.  
Section~\ref{twostacks} contains a description of the D3/D7/D7$'$ system,  
that is, two perpendicular stacks of the D7-branes that are parallel to the  
D3. This system has $\caln=1$ supersymmetry, and we discuss it in some  
detail. Section~\ref{sec:probe} presents a probe calculation of the scalar  
(meson) spectrum of the $\caln=1$ gauge theory. We provide the spectrum  
as an expansion in the number of D7-branes of the D3/D7 geometry.  
We then turn to the conifold picture by first reviewing  
the structure of the conifold as a K\"ahler manifold in section~\ref{con},  
and then describing a solution with D7-branes holomorphically embedded in  
the conifold in section~\ref{cond7}.  Section~\ref{gauge} contains  
some comments on the gauge theory side. We conclude  
in section~\ref{conclusions} by enumerating a number of open  
problems. Appendix~\ref{mam} contains some useful results  
pertaining to properties of the Monge-Amp\`ere equation.  
  
\section{D3/D7 Systems}\label{d3d7}  
  
In this section we review the basic setup for examining D3/D7 systems.  
The D3-brane geometry by itself breaks the 10-dimensional Poincar\'e  
invariance into manifest $\rm SO(3,1) \times SO(6)$.  This may be represented  
by a metric of the form  
\begin{equation}  
\eqlabel{metricSym}  
ds_{10}^2=h^{-1/2}(y_m)dx_\mu^2 + h^{1/2}(y_m)g_{mn}dy^mdy^n,  
\end{equation}  
where $x^\mu$ ($\mu=0,\ldots,3$) are flat coordinates on the longitudinal  
spacetime, and $y^m$ ($m=4,\ldots,9$) are coordinates transverse to the  
D3-brane.  The D7-branes are embedded  
in this 10 dimensional space such that they share the four longitudinal  
directions with the D3-branes while wrapping or spanning  
four out of the six transverse  
directions.  This embedding may be represented pictorially as  
\begin{center}  
\begin{tabular}{|c|c|c|c|c|c|c|c|c|c|c|}  
\hline  
&0&1&2&3&4&5&6&7&8&9\\  
\hline  
D$3$ &$-$&$-$&$-$&$-$&$\cdot$&$\cdot$&$\cdot$&$\cdot$&$\cdot$&$\cdot$\\  
\hline  
D$7$ &$-$&$-$&$-$&$-$&$-$&$-$&$-$&$-$&$\cdot$ &$\cdot$\\  
\hline  
\end{tabular}  
\end{center}  
for the case of a single stack of D7-branes.  
  
D7-brane configurations have been effectively studied in \cite{gsvy}  
in the context of cosmic strings,  
and D3/D7 systems have been examined in \cite{kehagias,pg}.  The starting  
point for exploration of D3/D7 geometries preserving some fraction of  
supersymmetries is the set of Killing spinor equations obtained from the IIB  
supersymmetry transformations \cite{Schwarz:qr}  
\begin{eqnarray}  
\d\l &=& i\g^M P_M\epsilon^*-\frac{i}{24}\g^{MNK}G_{MNK}\epsilon,\nonumber\\  
\d\psi_M &=& \left(D_{M}+\frac{i}{480}\gamma^{M_1 \ldots M_5} \gamma_M   
F_{M_1 \ldots M_5}\right)\epsilon-\frac{1}{96}\left(\gamma_M{}^{NPQ}  
-9\delta_M^N\gamma^{PQ}\right)G_{NPQ}\epsilon^*,\quad  
\label{dilatino}  
\end{eqnarray}  
where $D_M = \nabla_M - \frac{i}{2}Q_M$ and  
\begin{equation}  
P_M = \frac{i}{2}\frac{\partial_M \tau}{\tau_2},\qquad  
Q_M=-\fft{\partial_M\tau_1}{2\tau_2},  
\eqlabel{eq:pqdef}  
\end{equation}  
are the SL$(2,\mathbb R)$ scalar kinetic and composite connection terms for  
the dilaton/axion pair $\tau=\tau_1+i\tau_2$.  
  
The Killing spinor equations simplify in the absence of $G_3$ flux  
to yield  
\begin{eqnarray}  
\label{susy}  
\g^M P_M \epsilon^* &=& 0, \nonumber\\  
\left(D_{M}+\frac{i}{480}\gamma^{M_1M_2M_3M_4M_5}F_{M_1M_2M_3M_4M_5}\right)  
\epsilon&=&0.  
\end{eqnarray}  
Corresponding to the $4+6$ coordinate split of (\ref{metricSym}),  
we decompose the Dirac matrices according to $\gamma^{\mu}=(\Gamma^{\mu}  
\otimes 1)$, $\gamma^m=(\Gamma^5 \otimes \Gamma^m)$, where $\Gamma$  
represents either the SO$(1,3)$ or SO$(6)$ Dirac matrices as appropriate.  
This allows us to define the chirality matrices  
\begin{eqnarray}  
\gamma_{(4)}^5&=& i \gamma^0 \cdots \gamma^3=(\Gamma^5 \otimes 1),\nonumber\\  
\gamma_{(6)}^7&=& -i \gamma^4 \cdots \gamma^9=(1 \otimes \Gamma^7).  
\end{eqnarray}  
The complex IIB spinor $\epsilon$ has definite chirality, $\gamma^{11}  
\epsilon=\epsilon$.  Hence the four and six-dimensional chiralities are  
correlated, so that $\gamma_{(4)}^5\epsilon=\gamma_{(6)}^7\epsilon  
=s\epsilon$ where $s=\pm1$.  
In addition, the coordinate split also allows us to write the most  
general five form consistent with four-dimensional Poincar\'e symmetry  
and self-duality:  
\begin{equation}  
F_{M_1 M_2 M_3 M_4 M_5}=  
-\epsilon^{(6)}_{M_1 M_2 M_3 M_4 M_5 M_6}F^{M_6}  
+5\epsilon^{(4)}_{[M_1 M_2 M_3 M_4} F_{M_5]}^{\vphantom{(4)}},  
\end{equation}  
where $F^M$ is a function only of the $y^m$ coordinates, and is zero for  
$M=0\ldots3$.  To insure Poincar\'e symmetry we must also take $\tau$ to be  
a function only of the $y^m$ coordinates.  
  
Using the above, the Killing equations, (\ref{susy}), can now be written  
as  
\begin{eqnarray}  
P_m(1 \otimes \Gamma^m)\epsilon^*&=&0,\nonumber\\  
\partial_{\mu}\epsilon-\left( \frac{s}{8}\partial_n  
\log(h)+\frac{1}{2}F_n \right)(\Gamma_{\mu} \otimes \Gamma^n)\epsilon&=&0,  
\nonumber\\  
\nabla^{(6)}_m \epsilon-\frac {s}{2}F_m\epsilon+  
\left(\frac{1}{8}\partial_n  
\log(h)+ \frac{s}{2}F_n\right)(1 \otimes \Gamma_m{}^n)\epsilon  
-\frac{i}{2}Q_m\epsilon&=&0,  
\label{eq:splitkse}  
\end{eqnarray}  
where $\nabla^{(6)}$ is covariant with respect to the  
metric on the six dimensional space transverse to the D3 branes.  
Taking $\epsilon$ to be only a function of $y^m$, the  
second equation can immediately be solved by taking $\gamma^5_{(4)}\epsilon  
=\epsilon$ (so that $s=1$) and at the same time setting  
\begin{equation}  
F_m=-\frac{1}{4}\partial_m\log(h)=-\frac{\partial_m h}{4h}.  
\eqlabel{eq:5flux}  
\end{equation}  
This is a remarkable result, as it demonstrates that the usual relation  
between the D3-brane warp factor $h(y^m)$ and the five-form remains  
unchanged, even in the presence of D7-branes.  Note that the $s=1$  
condition on Killing spinors yields the familiar result that the  
D3-branes by themselves are half-BPS objects.  
  
Taking the warp factor as above, the Killing spinor equations  
(\ref{eq:splitkse}) reduce to  
\begin{equation}  
\gamma^m P_m \eta^* = 0,\qquad  
(\nabla_m - \frac{i}{2}Q_m)\eta = 0,  
\eqlabel{eq:d7kse}  
\end{equation}  
where $\eta=h^{1/8}\epsilon$.  These equations may be used to determine  
the structure of the D7-branes. Since the second equation above implies  
that $\eta$ is covariantly constant, the connection $\nabla_m$ is in  
U$(3)$, and the transverse space to the D3-branes is complex K\"ahler.  
The first of the above equations is a familiar one for D7-branes,  
namely $(\gamma^m\partial_m\tau)\eta^*=0$, and can be satisfied by taking  
$\tau$ to be holomorphic on the complex 3-fold.  This yields the further  
conditions $\gamma^m\eta^*=0$ on the Killing spinors whenever $\partial_m  
\tau\ne0$ (for the generic case), where $z_m$ $(m=1,2,3)$ is now a  
complex coordinate.  As a result, the D3/D7 system preserves $1/4$, $1/8$  
and $1/8$ of the IIB supersymmetries for $\tau$ depending holomorphically  
on one, two and three of the complex $z_m$ coordinates, respectively.  Note  
that there is no further reduction of supersymmetry for the last case, as  
the product of the three independent D7 projections gives precisely the  
D3 projection.  
  
The second equation in (\ref{eq:d7kse}) has a useful integrability condition,  
$R^{(6)}_{mn}=P_m^{\vphantom{*}}P^{*}_n +P^{*}_m P_n^{\vphantom{*}}$. This  
has a particularly nice form for K\"ahler geometry, namely  
$R^{(6)}_{m \bar{n}}=P_{m}^{\vphantom{*}}P^*_{\bar{n}}$,  
where $m$ and $\bar n$ denote complex indices.  Taking $\tau(z_m)$ to be  
holomorphic and using $P_m$ from (\ref{eq:pqdef}) results in  
\begin{equation}  
-\partial_{m}\partial_{\bar{n}}\log(\det(g^{(6)}))=  
-\partial_{m}\partial_{\bar{n}}\log(\tau-\tau^*),  
\end{equation}  
which may be integrated to give  
\be  
\eqlabel{mastereqn}  
\det(\partial_{m}\partial_{\bar{n}} \calf)=\Omega \bar{\Omega}\,\Im(\tau).  
\ee  
Here $\calf$ is the K\"ahler potential and $\Omega$ is an arbitrary  
holomorphic function accounting for the integration factors.  The same  
$\Omega$ is used for the holomorphic and antiholomorphic  
integration functions to insure that the K\"ahler potential is real.  
In cases with `radial' symmetry, Eq.~(\ref{mastereqn}) can be reduced to  
a real Monge Amp\`ere equation (see Appendix~\ref{mam} for further comments  
on this equation).  
  
Finally, returning to the D3-brane warp factor, the integrability condition  
combined with the Einstein equations gives the result that the function $h$  
is harmonic  
\begin{equation}  
\Box_{(6)} h=-(2\pi)^4N_c{\det(g^{(6)})}^{-1/2}  
\delta^6\left(y^m-y^m_{\D3}\right).  
\eqlabel{eq:6dhfunc}  
\end{equation}  
As a result, solving for the D3/D7 geometry can be reduced to several  
steps.  First, we take a (noncompact) Calabi-Yau background geometry,  
and then turn on the D7-branes by choosing a holomorphic $\tau$ with  
appropriate monodromies.  Then we must solve the complex Monge Amp\`ere  
equation (\ref{mastereqn}) for the K\"ahler potential, yielding the  
geometry of the six-dimensional transverse space.  Finally, to include the  
D3-branes, we must solve the harmonic function equation (\ref{eq:6dhfunc}).  
This gives the warp factor and corresponding five-form flux through  
(\ref{eq:5flux}).  Unlike for most intersecting brane configurations, at  
least in principle this procedure yields fully localized D3/D7 solutions,  
although the resulting equations are often quite intractable in practice.  
Note that the possibility of constructing localized intersecting solutions  
for the D3/D7 system is a feature shared with the D2/D6 system  
\cite{Cherkis:2002ir}.  
  
\subsection{D3 and a single stack of D7 in flat space}  
  
We now consider the simplest case where the six-dimensional transverse  
space to the D3-branes is flat before turning on any D7-branes. In particular  
\be  
ds_{6,\rm\;background} = dz_1d\bar z_1+dz_2d\bar z_2+dz_3d\bar z_3,\qquad  
\calf_{\rm flat,\;background}= z_1\bar z_1 + z_2\bar z_2 + z_3\bar z_3.  
\ee  
In the presence of D7-branes, the K\"ahler function may be obtained  
by solving (\ref{mastereqn}).  However, before doing so, we must  
understand the form of the dilaton/axion field, $\tau(z_1,z_2,z_3)$.  
  
For a single stack of D7-branes transverse to $z_3$, we may follow  
the procedure given in \cite{gsvy}, which we outline here.  
If one requires modular invariance for $\tau$, holomorphicity and  
the proper monodromy around the D7 branes, the basic solution can be  
written in terms of the modular invariant $j$-function as \cite{gsvy,pg}  
\be  
\eqlabel{odilaton}  
j(\tau)=f(z_3),  
\ee  
where for now $f$ is an arbitrary holomorphic function.  In \cite{gsvy}  
it was observed that poles in $f$ correspond to D7-branes, in the sense of  
decompactification at the cores of the branes (as will be indicated below).  
Solving (\ref{mastereqn}) with this form of the dilaton/axion is quite  
involved; hence we only consider the neighborhood close to one of the  
poles  
\begin{equation}  
\eqlabel{dilaton}  
j(\tau)=\frac{b}{(z_3-z_0)^{N_f}}.  
\end{equation}  
The r\^ole of the $j$-function in the  
above expression is to provide a map from the fundamental domain into  
the whole complex plane which is then related to $N_f$ coverings of  
the $z_3$-plane.  
  
We have limited ourselves to $z_3$ close to $z_0$ which requires that  
$\tau_2$ is large.  In this case the $j$ function can be approximated as  
$\exp(-2\pi i \tau)$, so that the above equation becomes  
\begin{equation}  
\tau(z_3)=i\left(\frac{\log(b)}{2\pi}-\frac{N_f}{2\pi}\log(z_3)\right),  
\qquad(z_3\ll1),  
\eqlabel{tauf}  
\end{equation}  
where we have shifted $z_3$ to remove $z_0$.  This is known as the  
``decompactification limit'' from F-theory because it corresponds to the  
modulus of the compactified torus going to infinity.  We will work  
in this limit for the remainder of the paper.  (Note that Eq.~(\ref{tauf})  
is no longer appropriate for $z_3\gtrsim1$, where instead $\tau$ approaches  
a constant value.)  It is also natural to  
consider scalings of $z_3$ so that $b$ may be removed as well.  Two  
scalings that are natural are  $b=1$ and $b=e^{2\pi}$,  
making the constant $\log(b)/{2\pi}$ either 0 or 1, respectively.  
  
In \cite{gsvy}, $\Omega$ is fixed by considering modular invariance of  
the metric, imposing regularity of the metric, and noting that  
$g_{3 \bar{3}}$ is proportional to $\tau_2$.  This gives  
\begin{equation}  
g_{3 \bar{3}} \sim \tau_2 |\eta(\tau)|^4|z_3|^{-N_f/6}.  
\end{equation}  
Near the core of the D7-brane, where $\tau_2$ is large, this has the  
simple form  
\begin{equation}  
g_{3 \bar{3}} \sim \tau_2.  
\eqlabel{eq:metcomp33}  
\end{equation}  
In particular, the additional factor $|\eta(\tau)|^4|z_3|^{-N_f/6}$ in  
the metric simply approaches a constant near the cores of the D7-branes.  
  
We now turn to the question of the K\"ahler potential.  We have chosen  
to stack the D7-branes together so that they are extended along $z_1$  
and $z_2$, and are transverse in $z_3$. Making use of the  
background symmetry, and noting that $\tau_2$ is a function of the  
real combination $z_3\bar{z}_3$ suggests that we take  
\be  
\calf_{D7} = z_1\bar z_1 + z_2\bar z_2 + f(z_3 \bar z_3).  
\ee  
The background K\"ahler potential is obtained from (\ref{mastereqn}) by taking  
$\Im\tau=1$ and $\Omega=1$, so that when $f=z_3\bar{z}_3$ we simply get  
flat space.  We now take $b=e^{2\pi}$ and turn on non-zero $N_f$, so that  
(\ref{mastereqn}) becomes  
\begin{equation}  
\partial_3\partial_{\bar 3}f(z_3\bar z_3)\equiv (z_3\bar z_3)f'' + f'  
= \left(1-\frac{N_f}{4\pi}\log(z_3 \bar z_3)\right).  
\end{equation}  
By making a change of variables, $y_3 = \log(z_3\bar z_3)$, we arrive at  
\be  
f''(y_3) = \left(1-\frac{N_f}{4\pi} y_3\right) e^{y_3}.  
\ee  
This may be straightforwardly integrated to obtain the K\"ahler function  
for a single stack of D7-branes located at $z_3=0$  
\be  
\calf_{\rm D7} = z_1\bar z_1 + z_2\bar z_2 + z_3\bar z_3  
-\frac{N_f}{4\pi}(z_3\bar z_3)(\log(z_3\bar z_3) -2)  
+ C_1\log(z_3\bar z_3)+ C_2.  
\ee  
The constants $C_1$ and $C_2$ do not affect the metric; we will  
therefore set them to zero (we may always add a holomorphic plus  
antiholomorphic function to a K\"ahler potential).  Of course, the  
transverse space metric is given by  
\be  
\eqlabel{d7metric}  
ds_6^2=dz_1d\bz_1+dz_2d\bz_2+  
\left(1-\frac{N_f}{4\pi }\log z_3\bz_3\right)dz_3d\bz_3,  
\ee  
and this could have been obtained directly from (\ref{eq:metcomp33})  
without even having to solve for the K\"ahler function.  Nevertheless,  
this exercise in obtaining $\calf$ by solving (\ref{mastereqn}) will prove  
essential when considering multiple stacks of D7-branes as well as  
D7-branes in other backgrounds.  
  
\subsection{The warp factor}\label{sec:warpfact}  
  
Now that we have found the D7-brane contribution to the transverse  
metric, (\ref{d7metric}), we may turn to the D3-brane warp factor.  
To determine the warp factor, we must solve the transverse Laplacian  
(\ref{eq:6dhfunc}).  For a single stack of D7-branes, we first rewrite  
the metric (\ref{d7metric}) in the generic form  
\begin{equation}  
ds_6^2=dz_1d\bar z_1+dz_2d\bar z_2+e^{\Psi(z_3,\bar z_3)}dz_3d\bar z_3,  
\end{equation}  
in which case the transverse Laplacian is more explicitly  
\begin{equation}  
\Box_{(6)} h = (\partial_{1}\partial_{\bar 1}+\partial_{2}\partial_{\bar 2}  
+e^{-\Psi}\partial_{3}\partial_{\bar 3}) h.  
\end{equation}  
This expression for the warp factor was examined in \cite{afm}, where it was  
solved to a first order approximation away from the D7-branes, where  
the metric is approximately flat, with a deficit angle.  
  
In the absence of D7-branes, it is natural to place the stack of D3-branes at  
the origin, $z_m=0$, in which case the SO(6) R-symmetry is preserved, so that  
$h=L^4/R^4$ where $R$ is the radial coordinate, $R^2=|z_1|^2+|z_2|^2+|z_3|^2$.  
In orienting the D7-branes to lie fully in the $z_1$ and $z_2$-planes, we  
preserve a natural $\rm SO(4)\times SO(2)\subset SO(6)$ subgroup.  Note,  
however, that separating the D3 and D7-branes in the $z_3$-plane will break  
the SO(2).  With this symmetry in mind, we introduce real coordinates  
\begin{equation}  
r^2=|z_1|^2+|z_2|^2,\qquad z_3=\rho_3 e^{i\phi_3},  
\end{equation}  
so that  
\begin{equation}  
\Box_{(6)} h = \left(\fft1{r^3}\partial_rr^3\partial_r+e^{-\Psi(\rho_3,\phi_3)}  
(\fft1{\rho_3}\partial_{\rho_3}\rho_3\partial_{\rho_3}+\fft1{\rho_3^2}  
\partial_{\phi_3}^2)\right)h(r,\rho_3,\phi_3).  
\eqlabel{eq:d3lap}  
\end{equation}  
If the D7-branes are separated by a distance $d$ along the real $z_3$-axis,  
the metric function $e^{\Psi}$ has the form  
\begin{equation}  
e^{\Psi}=\left(1-\fft{N_f}{4\pi}\log(\rho_3^2-2\rho_3 d\cos\phi_3+d^2)\right).  
\end{equation}  

Although no straightforward solutions to (\ref{eq:d3lap}) appear to exist  
in the general case, we may find approximate solutions in appropriate  
limits.  For example, when $d=0$ the D3-branes overlap the D7-branes, and  
the remaining SO(2) symmetry is restored.  In this case, the Laplacian  
reduces to one with only two radial variables, $\rho_3$ and $r$.  Furthermore,  
we may consider solving for the warp factor near the core of the D7-branes,  
where $e^\Psi\gg1$.  This is the decompactification region of the complex  
$z_3$-plane, and corresponds to the region where the backreaction is strong.

It can be checked  that the equation\footnote{We are thankful to  
A. Hashimoto and   
M. Mahato for discussions relevant to this topic.}  
\ref{eq:d3lap} allows separation of variables in the form:   
\begin{equation}  
h=f(r)g(\rho,\phi){\it Y}^{l}(S^3),   
\end{equation}  
where ${\it Y}^{l}(S^3)$ are spherical harmonics on $S^3$    
  
\begin{equation}  
{\nabla}^{i}{\nabla}_{i}{\it Y}^{l}=-l(l+2){\it Y}^{l}.  
\end{equation}  
The equation for the function $f(r)$ can be solved in terms of Bessel  
functions. The equation for $g$ can be further separated in variables,   
after Fourier expanding in an appropriate angular variable, resulting in  
an equation for the radial variable along. The latter is a more involved  
second order differential equation which we will not discuss here in detail.

\section{D3 and two stacks of D7-branes}\label{twostacks}  
  
The D3 geometry in the presence of a single stack of D7-branes is perhaps  
the most straightforward to investigate, as there is little ambiguity in  
how the D7-branes are turned on.  However, we are generally more interested  
in configurations admitting ${\cal N}=1$ duals with flavor.  This may  
be achieved by either starting with a transverse space to the D3-branes  
with reduced supersymmetry or by turning on multiple stacks of D7-branes.  
In this section we focus on the latter possibility by turning on two  
stacks of mutually perpendicular D7-branes.  
  
In particular, we can construct a solution representing two stacks of  
D7-branes by placing, as before, $N_3$ D7-branes perpendicular to the  
$z_3$-plane and another set of $N_2$ D7-branes transverse to the  
$z_2$-plane.  This configuration may be represented as follows  
\begin{center}  
\begin{tabular}{|c|c|c|c|c|c|c|c|c|c|c|}  
\hline  
&0&1&2&3&4&5&6&7&8&9\\  
\hline  
D$3$ &$-$&$-$&$-$&$-$&$\cdot$&$\cdot$&$\cdot$&$\cdot$&$\cdot$&$\cdot$\\  
\hline  
D$7$ &$-$&$-$&$-$&$-$&$-$&$-$&$-$&$-$&$\cdot$ &$\cdot$\\  
\hline  
D$7'$ &$-$&$-$&$-$&$-$&$-$&$-$&$\cdot$ &$\cdot$&$-$&$-$\\  
\hline  
\end{tabular}  
\end{center}  
Note that turning on a third stack of D7-branes transverse to $z_1$ would  
not further reduce any supersymmetry.  However for simplicity we avoid  
this case as well as other possible non-perpendicular configurations.  
  
In order to have two stacks of D7-branes, located at $z_2\to0$ and $z_3\to0$,  
we take $\tau(z_1,z_2,z_3)$ to have the appropriate monodromies around the  
respective cycles.  More precisely, near the D7-branes we have  
\begin{equation}  
\tau=i\left(1-\fft{N_2}{2\pi}\log z_2-\fft{N_3}{2\pi}\log z_3\right).  
\end{equation}  
As a result, it is natural to take an Ansatz for the K\"ahler function  
in the form  
\be  
\calf_{\rm D7/D7'}=z_1\bar z_1+ f(z_2\bar z_2, z_3\bar z_3).  
\ee  
Since both $\calf_{\rm D7/D7'}$ and $\Im\tau$ depend only on the magnitudes  
of the complex coordinates, we make a similar change of variables as in  
the single stack case  
\be  
y_2 = \log(z_2\bar z_2), \qquad y_3=\log(z_3\bar z_3),  
\ee  
so that the equation determining the K\"ahler potential, (\ref{mastereqn}),  
becomes  
\be  
\eqlabel{2stacks}  
f_{22}f_{33}-f_{23}{}^2=\left(1-\frac{N_2}{4\pi} y_2  
-\frac{N_3}{4\pi} y_3\right)e^{y_2+y_3},  
\ee  
where subscripts on $f$ denote differentiation with respect to the  
appropriate $y_2$ or $y_3$ variable.  
This is a non-linear second order differential equation, and to our  
knowledge there are no general methods for obtaining a solution.  
However, as we explain in Appendix~\ref{mam}, Eq.~(\ref{2stacks}) is a  
Monge-Amp\`ere equation in the two real variables $y_2$ and $y_3$. For  
such equations there are general existence theorems for solutions, given  
some fairly general boundary conditions (see Appendix~\ref{mam}).  
  
A crucial r\^ole in understanding the solutions of (\ref{2stacks}) is  
played by the choice of boundary conditions. In the following subsection  
we present a solution that is perturbative in the number of D7-branes.  
The boundary conditions that we impose imply that to lowest order the  
solution behave like two independent superposed stacks of D7-branes.  
  
\subsection{Perturbative solutions for two stacks of D7-branes}  
  
In order to develop a series solution to (\ref{2stacks}) where both  
$N_2$ and $N_3$ are small, we introduce an expansion parameter $\epsilon$  
and define  
\begin{equation}  
n_2=\epsilon\fft{N_2}{4\pi},\qquad n_3=\epsilon\fft{N_3}{4\pi}.  
\end{equation}  
We have also absorbed the factors of $4\pi$ to simplify the subsequent  
expressions.  Note that this expansion presupposes that both $N_2$ and  
$N_3$ are of the same order.  In this case, (\ref{2stacks}) takes the form  
\be  
f_{22}f_{33}-f_{23}{}^2=e^{y_2+y_3}(1-\epsilon(n_2y_2+n_3y_3)).  
\eqlabel{eq:2dete}  
\ee  
We now proceed to look for a perturbative solution of the form  
\be  
f(y_2,y_3)=f^{(0)}+ \sum_{n=1}^\infty \epsilon^n f^{(n)}.  
\ee  

The zeroth order solution corresponds to flat space with no  
D7-branes, and the corresponding part of the K\"ahler potential is  
simply given by  
\be  
f^{(0)}=e^{y_2}+e^{y_3}.  
\ee  
At the first order, we find  
\be  
e^{-y_2} f^{(1)}_{22}+e^{-y_3} f^{(1)}_{33} = -n_2y_2-n_3y_3.  
\ee  
In general, solutions to this equation are only determined up to  
choice of boundary conditions.  However, at the linear order, this can  
be viewed as a sum of two equations---one for each D7 stack.  Hence  
we choose a simple solution of the form  
\be  
f^{(1)}=- {n_2} e^{y_2} (y_2-2) -{n_3} e^{y_3} (y_3-2).  
\ee  
This may be viewed as defining our boundary conditions (at first order)  
corresponding to the desired two stack configuration.  
  
We now demonstrate how this series expansion may be developed to higher  
orders.  Although we have started with a non-linear equation, the  
perturbative expansion yields linear equations at each order.  In  
particular, for $n\ge2$, $f^{(n)}$ is determined by solving  
\begin{equation}  
e^{-y_2}f_{22}^{(n)}+e^{-y_3}f_{33}^{(n)}  
=-e^{-y_2-y_3}  
\sum_{i=1}^{n-1}\left(f_{22}^{(i)}f_{33}^{(n-i)}-f_{23}^{(i)}f_{23}^{(n-i)}  
\right),  
\qquad(n\ge2).  
\end{equation}  
Note that the linear operator $L=e^{-y_2}\partial_2^2+e^{-y_3}\partial_3^2$  
separates, allowing a solution to be obtained through separation of  
variables.  However we will not pursue this method here, but will  
instead develop the next few orders by inspection. The second order  
equation becomes  
\be  
e^{-y_2}f^{(2)}_{22}+e^{-y_3}f^{(2)}_{33}=-(n_2\,n_3) y_2\,y_3,  
\ee  
demonstrating that $y_2$ and $y_3$ become mixed at this order.  This  
equation has a solution of the form  
\be  
f^{(2)}=a e^{y_2}\,y_3(y_2-2) + b e^{y_3}\,y_2(y_3-2), \qquad  
a+b =-{n_2\,n_3}.  
\ee  
Of course, this solution is not unique.  However, it maintains the structure  
of the lower order terms, namely $f\sim e^{y_2}f_2(y_2,y_3)+e^{y_3}  
f_3(y_2,y_3)$.  In order to maintain symmetry between the two stacks,  
we take $a=b=-(n_2\,n_3)/2$.  Similarly, we find a solution to third order  
\be  
f^{(3)}=-\ft12n_2^2\,n_3^{\vphantom{2}}e^{y_2}\, y_3 (y_2^2-4y_2+6)  
-\ft12n_2^{\vphantom{2}}\,n_3^2e^{y_3}\, y_2 (y_3^2-4y_3+6),  
\ee  
which maintains interchange symmetry between $y_2\leftrightarrow y_3$  
(with corresponding $n_2\leftrightarrow n_3$).  Thus the above mentioned  
structure of $f$ as a sum of two independent terms persists to third  
order. It is worth remarking, however, that the solution necessarily has  
a different structure at fourth order in perturbation theory.  The reason  
is that precisely at this order the term $f_{23}{}^2$ starts contributing,  
and therefore breaks the symmetry of the first three orders.  
  
Collecting the above terms in the expansion, we find that the K\"ahler  
potential takes the form  
\bea  
\label{kpot2stacks}  
\calf_{\rm D7/D7'}&=&z_1\bar z_1+z_2\bar z_2+z_3\bar z_3  \nonumber \\  
&&-\frac{N_2}{4\pi}z_2\bz_2 \left(\log z_2\bz_2-2\right) -  
\frac{N_3}{4\pi}z_3\bz_3  \left(\log z_3\bz_3-2\right)\nonumber \\  
&&-\frac{N_2\,N_3}{32\pi^2}\Bigl[z_2\bz_2(\log z_2\bz_2-2)\log z_3\bz_3  
+ z_3\bz_3(\log z_3\bz_3-2)\log z_2\bz_2\Bigr] \nonumber \\  
&&-\frac{N_2^2\,N_3^{\vphantom{2}}}{128\pi^3}z_2\bz_2 \log z_3\bz_3  
\left(\log^2z_2\bz_2 -4\log z_2\bz_2 +6\right) \nonumber \\  
&&- \frac{N_2^{\vphantom{2}}\,N_3^2}{128\pi^3}z_3\bz_3 \log z_2\bz_2  
\left(\log^2z_3\bz_3 -4\log z_3\bz_3 +6\right)+\cdots.  
\eea  
Given the K\"ahler potential, we can easily read off the metric. We  
present a few entries; the rest can be recovered using the  
symmetries under $z_2$ and $z_3$:  
\bea  
\label{2stacksmixing}  
g_{2\bar{2}}&=&1-\frac{N_2}{4\pi}\log z_2\bz_2 -\frac{N_2\,  
N_3}{32\pi^2}\log z_2\bz_2 \log z_3\bz_3 -\frac{N_2^2\,N_3}{128\pi^3}  
\log^2 z_2\bz_2 \log z_3\bz_3+\cdots,\nonumber\\  
g_{3\bar{2}}&=&-\frac{N_2\,N_3}{32\pi^2}\left(\frac{z_2}{z_3}\,(\log  
z_2\bz_2-1) +\frac{\bz_3}{\bz_2}\,(\log z_3\bz_3-1)\right) \nonumber \\  
&&-\frac{N_2^2\,N_3^{\vphantom{2}}}{128\pi^3}\,\frac{z_2}{z_3}  
\left(\log^2 z_2\bz_2-2\log z_2\bz_2+2\right)  
-\frac{N_2^{\vphantom{2}}\,N_3^2}{128\pi^3}\,\frac{\bz_3}{\bz_2}  
\left(\log^2 z_3\bz_3-2\log z_3\bz_3+2\right)\nonumber\\  
&&+\cdots.  
\eea  
We see that, to first order in the number of D7-branes, the metric looks  
like a simple superposition of two stacks of D7.  This validates our  
implicit choice of boundary conditions imposed above.  However, already  
at second order we see a mixing between the $z_2$ and $z_3$ planes. Note  
that $g_{3\bar{2}}$ signals a highly nontrivial dependence on the  
coordinates $z_2$ and $z_3$ through  the presence of ratios  of  
complex coordinates.  
  
\subsection{A perturbation in the relative number of D7-branes}  
  
While in the above we have focused on $N_2$ and $N_3$ both of the  
same order, we may alternatively consider the case where one of the  
stacks of D7-branes is much smaller than the other.  To be concrete,  
we take $N_2\ll N_3$, and consider an expansion in terms of the  
small parameter $N_2/N_3$.  In this case, we define  
\begin{equation}  
n_2=\epsilon\fft{N_2}{4\pi},\qquad n_3=\fft{N_3}{4\pi},  
\end{equation}  
so that the determining equation, (\ref{2stacks}), becomes  
\begin{equation}  
f_{22}f_{33}-f_{23}{}^2=e^{y_2+y_3}((1-n_3y_3)-\epsilon n_2y_2).  
\end{equation}  
Clearly this equation may be expanded in the same manner as  
(\ref{eq:2dete}).  However, the source terms will be rearranged at  
the first two orders.  
  
At zeroth order, we take the exact solution for a single stack of $N_3$  
coincident D3-branes  
\begin{equation}  
f^{(0)}=e^{y_2}+e^{y_3}(1-n_3(y_3-2)).  
\end{equation}  
At first order, we find that $f^{(1)}$ must satisfy  
\begin{equation}  
e^{-y_2}(1-n_3y_3)f^{(1)}_{22}+e^{-y_3}f^{(1)}_{33}=-n_2y_2.  
\end{equation}  
This has a particularly simple solution of the form  
\begin{equation}  
f^{(1)}=-n_2e^{y_3}y_2.  
\end{equation}  
For higher orders, the perturbative expansion has the form  
\begin{equation}  
\left(e^{-y_2}\partial_2^2+\fft{e^{-y_3}}{1-n_3y_3}\partial_3^2\right)  
f^{(n)}=-\fft{e^{-y_2-y_3}}{1-n_3y_3}\sum_{i=1}^{n-1}  
\left(f_{22}^{(i)}f_{33}^{(n-i)}-f_{23}^{(i)}f_{23}^{(n-i)}\right),  
\qquad (n\ge2),  
\end{equation}  
which is again amenable to separation of variables.  
  
Collecting the terms in the K\"ahler potential computed above,  
we obtain  
\be  
\calf=z_1\bz_1 + z_2 \bz_2 +z_3 \bz_3 - \frac{N_3}{4\pi} z_3\bz_3(\log  
z_3\bz_3-2)   \nonumber \\  
- \frac{N_2}{4\pi} z_3\bz_3 \log z_2 \bz_2+{\cal O}(N_2^2).  
\ee  
This K\"ahler potential results in the following metric:  
\bea  
ds^2&=& dz_1 d\bz_1 +dz_2 d\bz_2  
+\left(1-\frac{N_3}{4\pi} \log z_3\bz_3-\frac{N_2}{4\pi} \log z_2\bz_2  
\right)dz_3 d\bz_3 \nonumber \\  
&&- \frac{N_2}{4\pi}\left(\frac{\bz_3}{\bz_2} dz_3 d\bz_2 +\frac{z_3}{z_2}  
dz_2 d\bz_3\right)+{\cal O}(N_2^2).  
\eea  
Note that the main effect of turning on a second stack of D7-branes  
transverse to $z_2$ is to change $g_{3\bar{3}}$ by an expected amount  
proportional to $\log z_2\bz_2$. What is slightly unexpected, however,  
is the appearance of non-diagonal terms $g_{2\bar{3}}$ at first order  
in the $N_2$ perturbation.  Nevertheless, this is consistent with the  
expansion of the previous subsection.  
  
\section{Probe analysis}\label{sec:probe}  
  
In the previous section, we have examined the K\"ahler structure of the  
background with two perpendicular stacks of D7-branes.  To proceed, we  
would now in principle need to solve for the D3-brane warp factor, much  
as we did for the single stack case in section \ref{sec:warpfact}.  
However, we may in fact already extract some $\caln=1$ flavor physics  
without full knowledge of the warp factor by considering a perpendicular  
D7$'$ probe in the backreacted D3/D7 background.  
  
Let us first reemphasize, however, that taking the F-theory expression for  
the dilaton, (\ref{dilaton}), implies that the dilaton/axion has a  
very different behavior in two regions of interest: near and far from  
the D7-branes. Near a stack of D7-branes we  
have an expression of the form (\ref{tauf}). However, far away from the  
stack (that is, for large values of $z$) the solution of  
(\ref{dilaton}) is roughly of the form  
\be  
\eqlabel{tauinfty}  
\tau=\tau_0+ \frac{c}{z^{N_f}}.  
\ee  
The above expression can be obtained from (\ref{dilaton}) by considering  
an expansion near $|z|\to \infty$. Note that the radius of  
convergence of the above expression excludes $z=0$. The leading term  
is nothing but a constant dilaton thus allowing us to recover flat space  
in this limit.  
  
The general solution to (\ref{mastereqn}) in the case of one stack of  
D7-branes implies that the metric entry $g_{z\bz}$ is simply  
proportional to $\tau_2$. The arbitrary function  
$\Omega$ of (\ref{mastereqn}) was fixed in \cite{gsvy} by imposing  
modular invariance and regularity of the metric, leading to  
\be  
g_{z\bz}=e^\Psi=\tau_2|\eta(\tau)|^4 |z|^{-N_f/6}.  
\ee  
In the large $|z|$ limit that we are interested in in this section, the  
metric becomes that of flat space with a deficit angle. Namely, for  
large $|z|$ the dilaton is constant to first order and therefore we  
have for the relevant part of the metric (defining $z=re^{i\phi}$)  
\be  
ds_\perp^2=r^{-N_f/6}\left(dr^2 +r^2  
d\phi^2\right)=\frac{1}{(1-\frac{N_f}{12})^2}\bigg[d\rho^2+\rho^2  
(1-\frac{N_f}{12})^2 d\phi^2\bigg],  
\ee  
where $\rho=r^{1-N_f/12}$. Thus the effect of the D7 is to create  
a deficit angle at asymptotic  infinity. Due to this deficit angle  
we are forced to consider $N_f\le12$ (where $\rho=\log r$ must be  
taken for $N_f=12$)%
\footnote{Actually, two asymptotic geometries  
with $N_f=12$ may be joined to provide the F-theory fibration of K3 over  
a compact base.  However, we do not consider this case.}.  

To be specific, we take the orientation of the probe D7$'$ brane as indicated  
in the table below  
\begin{center}  
\begin{tabular}{|c|c|c|c|c|c|c|c|c|c|c|}  
\hline  
&0&1&2&3&4&5&6&7&8&9\\  
\hline  
D$7$ &$-$&$-$&$-$&$-$&$-$&$-$&$-$&$-$&$\cdot$ &$\cdot$\\  
\hline  
D$7'$ &$-$&$-$&$-$&$-$&$\cdot$&$\cdot$&$-$ &$-$&$-$&$-$\\  
\hline  
\end{tabular}  
\end{center}  
The D3/D7 background is  
\bea  
ds^2&=&h^{-1/2} ds_4^2 + h^{1/2} (dz_1d\bar z_1 + dz_2 d\bar z_2 +  
e^{\Psi}dz_3 d\bar z_3),\nonumber\\  
C_{0123}&=&-h^{-1},\qquad  
\Box_{(6)} h=-(2\pi)^4 N_c e^{-\Psi}\delta^6(z).  
\eea  
In the static gauge, with an Ansatz such that angular coordinate  
$\phi_1$ in the $z_1$-plane is constant and the radial coordinate  
has a profile $\rho_1=\rho_1(\rho_3)$, the pull-back metric on the D7$'$  
probe becomes  
\be  
{}^*g_{ab} d\sigma^a d\sigma^b=h^{-1/2}ds_4^2+h^{1/2}  
(dz_2 d\bar z_2 + (e^\Psi+\dot\rho_1{}^2) d\rho_3^2 +  
e^\Psi \rho_3^2 d\phi_3{}^2),  
\ee  
where $\dot \rho_1=\frac{d}{d\rho_3}\rho_1$. The action of the D7$'$ probe  
\be  
S=-\mu_7\int d^8 \sigma e^{-\Phi}\sqrt{-\det({}^*g+2\pi \a'{\bf F})}+  
\mu_7\int {}^*C_{(8)} + \frac{\mu_7}{2(2\pi\a')^2}  
\int^{}*C_{(4)}\wedge {\bf F}\wedge{\bf F}  
\ee  
becomes (up to some infinite volume factor $2\pi Vol(R^6)$)  
\be  
S=-\mu_7\int d\rho_3 e^{-\Phi}e^{\Psi/2}\rho_3\sqrt{e^{\Psi}+\dot\rho_1{}^2}.  
\ee  
The equation of motion for $\rho_1$ determines the profile of the  
probe brane  
\be  
\frac{d}{d\rho_3}\left(e^{-\Phi}e^{\Psi/2}\rho_3  
\frac{\dot\rho_1}{\sqrt{e^{\Psi}+  
\dot\rho_1{}^2}}\right)=0,  
\ee  
where in the asymptotic region $e^\Psi=\rho_3^{-N_f/6}$ and  
$e^{\Phi}={\rm const}$%
\footnote{From $\tau=\tau_{\infty}+\frac c{(\rho_3)^{N_f}}  
e^{-iN_f\phi_3}$, we extract the dilaton  
$e^{-\Phi}=\Im\tau=\Im\tau_{\infty}-\frac c{\rho^{N_f}} \sin(N_f\phi_3)$.  
Furthermore, upon integration over $\phi_3$, we remain only with the  
constant dilaton factor in the effective action of the D7$'$ probe.}.  
The most general solution in the asymptotic region is given by  
\bea  
\rho_1&=&\int d\rho_3\sqrt\frac{C_1}{\rho_3^2-C_1\rho_3^{N_f/6}}\nonumber\\  
&=&\frac{\sqrt{C_1}}{N-12}\left((N-12)\log\rho_3  
- 12\log\left(1+\sqrt{1-C_1\rho_3^{N_f/6-2}}\right)\right)+C_2  
\eea  
where $C_1, C_2$ are integration constants.  
  
Since $\rho_1=\rm const.$ is a solution, the probe can be placed  
at a fixed position in the $z_1$ plane and therefore  
the probe has a flat profile. A consequence of this fact will be the  
absence of a quark condensate, {\it i.e.} $\langle Q\bar Q\rangle=0$.  
  
\subsection{The spectrum of the scalar fields}  
  
By expanding the probe action in fluctuations around the flat profile solution  
$\rho_1=d$, $\phi_1=\rm const.$, which is a solution even in the presence of  
a running dilaton, and identifying the transverse fluctuations $X,Y$  
with the scalar fields of the dual gauge theory, we extract their action  
(to leading order) as  
\be  
S=-\mu_7\int\sqrt{\det {}^*g}\,e^{-\Phi}(1+{}^*g^{ab}  
(\partial_a X\partial_b X+\partial_a Y\partial_b Y)),  
\ee  
where  
\bea  
\sqrt{\det{}^*g}&=&\rho_2\rho_3e^\Psi,\qquad  
e^{\Psi}=1-\frac{N_f}{2\pi}\log(\rho_3),\nonumber\\  
e^{-\Phi}&=&1-\frac{N_f}{2\pi}\log(\rho_3).  
\eea  
The equation of motion for either of the two scalar fields (collectively  
denoted by $X$) is  
\bea  
h\eta^{\mu\nu}  
\partial_\mu\partial_\nu X + \bigg[\frac{1}{\rho_2}\partial_{\rho_2}  
(\rho_2\partial_{\rho_2} X) + \frac{1}{\rho_2^2}\partial_{\phi_2}^2 X\bigg]  
&&\nonumber\\  
&&\kern-5cm  
+e^{-\Psi}\bigg[ \frac{1}{\rho_3}\partial_{\rho_3}  
(\rho_3\partial_{\rho_3} X) + \frac{1}{\rho_3^2}\partial_{\phi_3}^2 X -  
\partial_{\rho_3}\Phi\partial_{\rho_3}X \bigg]=0,  
\label{sfeom}  
\eea  
where we recall that the warp factor $h$ obeys  
\be  
(\partial_{1}\partial_{\bar 1} + \partial_2\partial_{\bar 2}) h  
+e^{-\Psi}\left(\frac{1}{\rho_3}\partial_{\rho_3} \rho_3\partial_{\rho_3} +  
\frac 1{\rho_3^2} \partial_{\phi_3}^2\right) h = -(2\pi)^4N_c e^{-\Psi}  
\delta^6(z).  
\ee  
When substituting the warp factor into (\ref{sfeom}) we evaluate it  
at a fixed position in the $z_1$-plane, namely  
$h=h(\rho_1=d,\phi_1={\rm const.},\rho_2,\rho_3)$.  
Even though the warp function equation cannot be solved exactly, we can  
attempt to solve it perturbatively in $N_f$, the number of D7-branes.  
With the change of variable $e^{\Psi/2}\rho_3 = r_3$, the metric in the  
$z_3$-plane becomes  
\bea  
\rho_3&\approx&r_3\left(1+\frac{N_f}{4\pi}\log r_3\right),\nonumber\\  
ds^2&=&e^\Psi \left(\frac{d\,\rho_3}{d \,r_3}\right)^2 d\,r_3^2  
+r_3^2 d\,\phi_3^2\nonumber\\  
&\approx&\left(1+\frac{N_f}{2\pi}\right)d\,r_3^2+r_3^2 d\,\phi_3^2,  
\eea  
and the corresponding Laplacian can be approximated by  
\be  
e^{-\Psi}\partial_3\partial_{\bar 3}\approx  
\left(1-\frac{N_f}{2\pi}\right)  
\frac{1}{r_3}\partial_{r_3}(r_3\partial_{r_3})+\frac{1}{r_3^2}  
\partial_{\phi_3}^2.  
\ee  
The effect of the D7-branes on the background, to first order, is to  
generate a conical deficit in the otherwise flat $z_3$ directions.  
  
One way to proceed is to take full advantage of the observation that,  
to first order in $N_f$, the geometry of the six-dimensional  
space transverse to the D3-branes is (conformally) flat, with a conical  
deficit in the $z_3$-plane. The warp factor is then simply the Green's  
function \cite{sh} $G(\vec z, 0)$ associated  
with this six-dimensional geometry:  
\bea  
ds_6^3&\approx& dz_1\bar dz_1 + dz_2d\bar z_2+  
dr_3^2 + r_3^2 d\varphi^2,\nonumber\\  
\varphi&\sim&\varphi  
+ \left(1-\frac{N_f}{4\pi}\right) 2n\pi\equiv\varphi + 2\nu \pi,\nonumber\\  
G(r_3^2,\rho_2^2+\rho_1^2,\varphi)  
&=&\frac{1}{\rho_2^2+\rho_1^2+r_3^2}\bigg[1\nonumber\\  
&&-2 \bigg(\ \frac{\sin(\varphi-\nu\pi)}{\sqrt{2\cos(2\varphi-2\nu\pi)-2}}  
\rm{arctanh}\frac{2(1+\cos(\varphi-\nu\pi))}{\sqrt{2\cos(2\varphi  
-2\nu\pi)-2}}\nonumber\\  
&&\hphantom{2\bigg(}
-\frac{\sin(\varphi+\nu\pi)}{\sqrt{2\cos(2\varphi+2\nu\pi)-2}}  
\rm{arctanh}\frac{2(1+\cos(\varphi+\nu\pi))}{\sqrt{2\cos(2\varphi  
+2\nu\pi)-2}}  
\bigg)\bigg],\nonumber\\  
h&\approx&L^4 G(r_3^2,\rho_2^2+d^2)\approx L^4\left(1-\frac{N_f}2\right)  
\frac{1}{r_3^2+\rho_2^2+d^2}.  
\eea  
Thus to first order in $N_f$ the effect of the D7-branes on the warp factor  
is a mere rescaling with $(1-N_f/2)$. This will carry through when  
investigating the meson spectrum.  
  
As we discussed earlier, the equation of motion for the two scalar fields  
corresponding to fluctuations in the transverse directions $z_1$,  
\be  
X(r_3,\varphi, \rho_2^2+d^2)\sim X(r_3,\varphi+2\nu\pi, \rho_2^2+d^2),  
\ee  
can be written as  
\be  
(\Box_4 + M^2 h)X=0.  
\ee  
where we have Fourier expanded the scalar fluctuations in the directions  
common with the D3-branes, $X=\int d^4 k e^{ik^\mu x_\mu} X(k)$, and we  
have identified the Casimir invariant $k^2$ with the four-dimensional mass  
$k^2=-M^2$.  We continue to expand $X$ into Fourier modes, according to the  
identifications of the wedge geometry  
\be  
X=\sum_{m,n} X_{m,n}(\rho_2,r_3)e^{im\phi_2}  
e^{i\nu n \varphi}.  
\ee  
By substituting into the equation of motion, we find  
\be  
\left(\frac{1}{\rho_2}\partial_{\rho_2} \rho_2\partial_{\rho_2}+\frac 1{r_3}  
\partial_{r_3} r_3\partial_{r_3} +\frac{N_f}{2\pi r_3}\partial_{r_3}  
- \frac{m^2}{\rho_2^2}-\frac{n^2\nu^2}{r_3^2}  
-M^2 h\right) X_{m,n}=0.  
\ee  
The next step is to make yet another change of variables,  
$r_3=R\cos\alpha$, $\rho_2 = R\sin\alpha$, and to further  
expand $X_{m,n}(R,\alpha)$ into  
the eigenfunctions of the differential operator  
\bea  
{\cal L}^2&=&\frac{1}{\sin\alpha\cos\alpha}\partial_{\alpha}(\sin\alpha  
\cos\alpha\partial_\alpha)-\frac{n^2\nu^2}{\cos^2\alpha}  
-\frac{m^2}{\sin^2\alpha},\nonumber\\  
{\cal L}^2 \xi_\l(\alpha)&=&\l\,\xi_\l(\alpha),\nonumber\\  
X(R,\alpha)&=&X(R)\xi_\l(\alpha),  
\eea  
where $\lambda=-l(l+2)$ and $l=|m|+|n|\nu$.  
We have thus reduced the equation of motion of the scalar fluctuations  
to an ordinary differential equation for $X(R)$:  
\be  
\bigg(\frac{1}{R^3}\partial_{R}R^3\partial_R +\frac 1{R^2}\l+  
\frac{N_f}{2\pi R}\partial_R + M^2 h \bigg)X(R)=0.  
\ee  
The square normalizable solution is  
\bea  
X(R)&=&R^{-1-\frac{N_f}{4\pi}+b}(R^2+d^2)^{-a}  
{}_2F_1(-a,\frac 12+b-c;-2a;R^2+d^2),\nonumber\\  
a&=&-\frac12+\frac{\sqrt{1+M^2(1-N_f/2)}}{2},\qquad  
b=\sqrt{(l+1)^2-N_f/(2\pi)+N_f^2/(4\pi)^2},\nonumber\\  
c&=&\frac{\sqrt{4+4M^2-2\pi M^2 N_f}}{\pi},  
\eea  
where we have to impose the additional constraint that the hypergeometric  
series terminates:  
\be  
\frac 12+b-c=-N,  
\ee  
with $N$ a positive integer.  
To first order in the number of background D7-branes, the mass spectrum  
of the D7$'$ probe scalar fluctuations is  
\bea  
M^2&=&4(N+|m|+|n|+2)(N+|m|+|n|+1)\nonumber\\&&+  
N_f\bigg(2(N+|m|+|n|+2)(N+|m|+|n|+1)\nonumber\\&&  
-\frac{(3+2(|m|+|n|)+2N)(n^2+|n|+|mn|-1)}  
{\pi(|m|+|n|+1)}\bigg).  
\eea  
Notice that by setting $N_f$ to zero we recover the spectrum of   
the scalar fluctuations  
of a  D7$'$ probe in the AdS$_5\times S^5$ derived in \cite{martin1}.   
Holographically, the scalar fluctuations in \cite{martin1} correspond to   
the spinless mesons  
of an ${\cal N}=2$ SYM gauge theory, whereas in our case, the mesons belong  
to a theory with reduced supersymmetry (${\cal N}=1$) and flavor.  
  
\section{Review of the conifold as a K\"ahler manifold}\label{con}  
  
After exploring the basic D3/D7 and D3/D7/D7$'$ systems, we now turn to  
the conifold.  Here, we review the construction of the Ricci-flat metric  
on the conifold. We anticipate that the level of detail included here will  
provide a better understanding of the choices that we make in the next  
section when considering the inclusion of D7-branes on the conifold.  
We follow the presentation of \cite{candelas} but attempt to make some of  
the computations more explicit.  
  
We begin by noting that the conifold is defined as the following quadric  
in $\mathbb{C}^4$:  
\be  
\eqlabel{wcoord}  
\sum\limits_{a=1}^4(w^a)^2=0,  
\ee  
where $w^a$ are complex coordinates. It is convenient to define a  
radial coordinate as  
\be  
\sum\limits_{a=1}^4|w^a|^2=r^2.  
\ee  
The symmetries and general form of the conifold can be made manifest by  
writing the defining equation as:  
\begin{equation}  
\eqlabel{defining}  
\mbox{det}\ \W=0, \qquad\hbox{\it i.e.}\qquad z_1z_2-z_3z_4=0,  
\ee  
where  
\be  
\eqlabel{zcoord}  
\W=\fft12\begin{pmatrix}  
w_3+iw_4&w_1-iw_2\\  
w_1+iw_2&-w_3+iw_4  
\end{pmatrix}  
\equiv  
\begin{pmatrix}  
z_1&z_3\\  
z_4&z_2  
\end{pmatrix}.  
\end{equation}  
In this case, the radial coordinate is given by  
\begin{equation}  
r^2=\mbox{tr} (\W^{\dagger} \W).  
\eqlabel{eq:rcoord}  
\end{equation}  

We now clarify the K\"ahler structure, with an eye toward generalizations that  
include non Ricci-flat metrics. If a metric on the conifold is K\"ahler it can be written as  
\be  
g_{\m\bar{\n}}=\del_\m\del_{\bar{\n}} \calf,  
\ee  
where $\calf$ is the K\"ahler potential. A K\"ahler potential  
invariant under $\rm SU(2)\times SU(2)$ will be a function of $r^2$ only.  
For such K\"ahler potentials, the metric can be written as  
\be  
\eqlabel{metric}  
g_{\m\bar{\n}}=\calf'\, \del_\m\del_{\bar{\n}} r^2 + \calf''\del_\m  
r^2\, \del_{\bar{\n}} r^2,  
\ee  
where prime means differentiation with respect to $r^2$.  
In terms of $\calw$ the metric takes the form  
\be  
ds^2=\calf'\mbox{tr}\left(d\calw^{\dagger}d\calw\right)+\calf''  
|\mbox{tr}\calw^{\dagger}d\calw|^2.  
\ee  
The Ricci tensor for a K\"ahler manifold is given by  
\be  
R_{\m\bn}=-\del_\m \del_{\bn}\, \log g,  
\ee  
where $g=\det g_{\m\bn}$. The Ricci flatness condition is easily  
obtained from computing the determinant of the metric  
(\ref{metric}). In particular  
\be  
\eqlabel{det}  
g=\det g_{\m\bn}=  
\frac{1}{|w_4|^2}\left(r^2(\calf')^3+r^4(\calf')^2\calf''\right).  
\ee  
As not all coordinates are independent, one of the coordinates, say  
$w_4$, may be eliminated in favor of the  
other three complex coordinates, {\it i.e.}~$w_4^2=-w_1^2-w_2^2-w_3^2$.  
  
The question of finding the Ricci flatness condition can now be  
rephrased as finding the K\"ahler potential such that (\ref{det}) is a  
product of a holomorphic and antiholomorphic function  
({\it e.g.}~$|\omega_4|^2$).  
The simplest way to achieve this is by introducing  
$\g=r^2\calf'$. The determinant can now be written as  
\be  
g=\frac1{2|w_4|^2r^2}(\g^3)'.  
\ee  
It now follows that choosing $\g \propto r^{4/3}$ will remove all $r$  
dependence. This yields the K\"ahler potential  
\be  
\eqlabel{kpcon}  
\calf=(r^2)^{2/3}=(w_1\bw_1+w_2\bw_2+w_3\bw_3+w_4\bw_4)^{2/3}.  
\ee  

After introducing  
$\r=\sqrt{\frac{3}{2}}r^{2/3}$, we obtain the standard Ricci-flat  
metric on the conifold:  
\be  
\eqlabel{conmetric}  
ds^2=d\r^2+\r^2\left(\frac16\sum\limits_{i=1}^2(d\theta_i^2  
+\sin\theta_i^2d\p_i^2)+\frac19(d\psi  
+ \sum\limits_{i=1}^2\cos\theta_id\p_i)^2\right).  
\ee  
The angular variables arise from the natural $\rm SU(2)\times SU(2)$  
symmetries of the conifold. Furthermore, they allow us to write the  
general solution of the defining equation (\ref{defining}) as  
\be  
\eqlabel{wparametrization}  
\calw= r L_1^{\vphantom{\dagger}} Z_0^{\vphantom{\dagger}} L_2^{\dagger},  
\ee  
where the $L_i$'s are $SU(2)$ matrices and $Z_0$ is a particular solution to  
the defining relation; for example  
\be  
\eqlabel{su2}  
L_j=\begin{pmatrix}  
\cos\frac12\theta_j\,e^{\frac{i}{2}(\psi_j+\phi_j)}&  
-\sin\frac12\theta_j\,e^{-\frac{i}{2}(\psi_j-\phi_j)}\\  
\sin\frac12\theta_j\,e^{\frac{i}{2}(\psi_j-\phi_j)}&  
\hphantom{-}\cos\frac12\theta_j\,e^{-\frac{i}{2}(\psi_j+\phi_j)}\\  
\end{pmatrix}, \qquad  
Z_0=\begin{pmatrix}  
0&1\\  
0&0\\  
\end{pmatrix}.  
\ee  
Note that, as a result of the coset structure of the conifold,  only the combination $\psi=\psi_1+\psi_2$ appears in the metric (\ref{conmetric}).  
  
\section{D3/D7 on the conifold}\label{cond7}  
  
The idea of considering a stack of D3-branes at the apex of the  
conifold was originally considered in \cite{kw} as a way to obtain  
$\caln=1$ superconformal gauge theories. This model is very attractive  
from a variety of viewpoints, and has been subsequently developed into  
one of the trademark examples of the gauge/gravity correspondence \cite{ks}.  
  
In this section we consider adding D7-branes to the Klebanov-Witten  
background. We explicitly develop two ways of adding D7-branes which  
are characterized by the method of embedding of the D7-branes.  
  
Our approach to including backreacted D7-branes in the geometry  
of the conifold is rooted in preserving the K\"ahler structure.  The  
supersymmetry and integrability conditions give, as before, the same  
condition presented in (\ref{mastereqn}).  
  
\subsection{The $w_4=0$ embedding }  
  
Given the defining equation of the conifold, (\ref{wcoord}), a natural  
embedding is to place the D7-branes at one of the $w$'s equal to a  
constant. Throughout  
this section we consider the embedding where the D7-branes are defined  
by $w_4=0$. Note, however, that we can use the SO$(4)$ symmetry of the  
conifold to substitute $w_4$ for any other $w_i$. The  
defining equation of the conifold and the definition of its radial  
coordinate preserve the SO$(4)$ symmetry. This symmetry prevents us  
from finding a simple K\"ahler potential corresponding to the inclusion of  
the D7-branes. In particular, it can be shown that there are no  
modifications of the K\"ahler potential (\ref{kpcon}) of the form  
$\calf=(r^2f(w_4,\bw_4)+h(w_4,\bw_4))^{2/3}$ for arbitrary functions  
$f$ and $h$.  
  
Being confronted with this fact rules out the most natural  
modification to $\calf$. Thus, taking into account the symmetries of the  
conifold, we are led to consider solutions in two variables of  
the form $\calf=\calf(s,t)$ where  
\be  
s=w_1\bw_1+w_2\bw_2+w_3\bw_3, \quad  
t=\sqrt{w_1^2+w_2^2+w_3^2}\sqrt{\bw_1^2+\bw_2^2+\bw_3^2}.  
\eqlabel{eq:stdef}  
\ee  
We now seek a solution to Eq.~(\ref{mastereqn})  
with $\Im\tau=1-n_f\log t$, where $n_f=N_f/(4\pi)$:  
\be  
\det(\partial_{m}\partial_{\bar{n}}\calf)=\Omega\bar\Omega\,\Im \tau  
=\left(\frac{2}{3}\right)^4\frac{1-n_f\log t}{t}.  
\ee  
The denominator and constant factor on the right hand side are to ensure  
that the Ricci-flat solution $\calf=(r^2)^{2/3}$ works for $n_f=0$.  
With the K\"ahler potential taken to be a function of both $s$ and $t$,  
this equation becomes  
\bea  
\calf_s^3+\calf_s^2(s(\calf_{ss}+\calf_{tt}+t^{-1}\calf_t)  
+2t\calf_{st})\qquad&&  
\nonumber\\  
+(s^2-t^2)\calf_s(\calf_{ss}(\calf_{tt}+t^{-1}\calf_t)-\calf_{st}^2)  
&=&\left(\frac23\right)^4(1-n_f\log t).  
\label{eq:conieqn}  
\eea  
We look for a linearized solution of the form  
\be  
\calf(s,t)=(s+t)^{2/3}\bigg[1+n_f\calf^{(1)}+n_f^2\calf^{(2)}+n_f^3\calf^{(3)}  
+\cdots \bigg].  
\ee  
Note that the zeroth order solution naturally corresponds to the conifold  
in the absence of D7-branes.  
  
We now observe that the linearization of (\ref{eq:conieqn}) splits the  
perturbation problem into an infinite set of equations that can be  
schematically written as  
\bea  
L \, \calf^{(1)}&=& \frac{4}{3}\log t, \nonumber \\  
L\, \calf^{(2)}&=& S_2\big[\calf^{(1)}\big]\nonumber \\  
\ldots &= & \ldots \nonumber \\  
L\, \calf^{(n)}&=& S_n\big[\calf^{(1)},\ldots, \calf^{(n-1)}\big],  
\eea  
where $S_i$ are functions that can be explicitly determined. The most  
important property is that at any order $i$, the function $S_i$ depends on already determined functions of lower order in the perturbation  
expansion. In addition, the same second order differential operator  
\bea  
\label{L}  
L&=&(3s^2+2st-2t^2)\partial^2_{s} + (2st+t^2)\partial^2_{t}+  
(2st+4t^2)\partial_{s}\partial_t\nonumber\\  
&&+(9s+6t)\partial_s + (2s+5t)\partial_t + 4  
\nonumber\\  
&=&(s+t)^{-2/3}\begin{pmatrix}\partial_s&\partial_t\end{pmatrix}  
(s+t)^{2/3}\begin{pmatrix}3s^2+2st-2t^2&st+2t^2\cr  
st+2t^2&2st+t^2\end{pmatrix}  
\begin{pmatrix}\partial_s\cr\partial_t\end{pmatrix}+4\quad  
\eea  
appears at every level in the perturbation expansion.  
Given that the operator $L$ is linear, and given the general form of the  
sources, we conclude that the above system is consistent.  Moreover,  
one can, in principle, find a solution for any given set of boundary  
conditions.  
  
In general, the functions $\calf^{(n)}$ depend on both $s$ and $t$. However,  
it turns out that for $\calf^{(1)}$ we can find a solution of the form  
\be  
\calf^{(1)}=\frac{1}{3}(\log t-1).  
\ee  
At second order in perturbation theory the situation is radically  
different, as there is no solution depending only on $t$.  The equation  
at second order is  
\be  
L \calf^{(2)}=\left(\frac32\right)^4\big[\frac12\frac{s+t}{t}-\log^2 t\big].  
\eqlabel{eq:c2order}  
\ee  
To solve the second (and higher) order equation in general, we would like  
to find a Green's function  
for the linear operator $L$.  To do so, it is convenient to  
perform the following change of variables:  
\be  
s=\rho \cos^2\theta, \qquad t =\rho\sin^2\theta.  
\ee  
In this case the operator $L$ in (\ref{L}) becomes  
\be  
L=3\r^2\del^2_\r+11 \r \del_\r +4 + (1-\ft12\sec^2\theta)\del^2_\theta  
+ \ft12(\cot\theta -\tan\theta(3+\sec^2 \theta))\del_\theta.  
\ee  

A general solution to the homogeneous equation  
can be constructed using separation of variables. Namely we take  
\be  
\calf^{(2)}(\r,\theta)=R(\r)\chi(\theta).  
\ee  
The $\theta$ equation, $L\chi(\theta)=\lambda\chi(\theta)$, is solved  
by Legendre polynomials  
\be  
\chi=P_n(-3 +4\cos^2 \theta), \quad n=\frac{\sqrt{5-\lambda}-1}{2}.  
\ee  
Note that, because of the way $s$ and $t$ are defined in (\ref{eq:stdef})  
the angular variable is restricted to $\fft12\le\cos^2\theta\le1$.  Thus  
the argument of the Legendre polynomial lies in the range $[-1,1]$.  
The remaining $R(\rho)$ equation is then homogeneous in powers of $\rho$;  
hence it may be solved by taking $R\sim\rho^\lambda$.  This gives the  
general solution to the homogeneous equation of the form:  
\be  
\calf^{(2)}=\sum\limits_{n=0}^\infty c_n^\pm \rho^{ -\frac{2}{3}(2\pm  
  \sqrt{1+3n(n+1)})} P_n(-3+4\cos^2\theta),  
\eqlabel{eq:cfexpand}  
\ee  
where the $c_n^\pm$ are coefficients that must be determined by boundary  
conditions.  Of course, to obtain the actual solution to the inhomogeneous  
equation (\ref{eq:c2order}), we would have to proceed with an actual  
construction of the Green's function.  Nevertheless, even without doing  
so, we can discern some general features of the solution.  For example,  
the r\^ole of $n$ in (\ref{eq:cfexpand}) can be understood as follows.  
Given that $\rho=s+t$ is the distance from the D3-brane, it is natural to  
conclude that higher terms in $n$ correspond to deforming the gauge  
theory by operators with higher conformal dimension.  
  
{}From the K\"ahler potential we can easily derive the metric to first  
order in the number of D7-branes, $N_f$.  
``Turning on'' the D7-branes alters the $g_{m\,\bar n}^0$ conifold geometry  
in two ways:  
\be  
g_{m\, \bar n} = g_{m\, \bar n}^{0} \left(1+\fft{N_f}3 (\log t - 1)\right)  
- \frac {N_f}3 r^{4/3} \partial_m w^4 \partial_{\bar n} \bar w^4.  
\eqlabel{defconmetr}  
\ee  
First, there is the rescaling of the background metric with a factor  
$(1+N_f (\log t - 1)/3)$. Second, we notice the presence of some  
off-diagonal metric components due to the last term \,in (\ref{defconmetr}).  
More explicitly, to first order in $N_f$, the correction to the  
six-dimensional conifold geometry reads  
\bea  
ds^2&=&\left(1+\fft{N_f}3(2 \log |w_4|-1)\right)(ds^0)^2\nonumber\\  
&&-\frac{N_f}{3}r^{4/3}\bigg[\fft12(1-\cos\theta_1\cos\theta_2  
-2\sin\theta_1\sin\theta_2\cos(\phi_1+\phi_2))  
dr^2\nonumber\\  
&&+ \frac 12 r(\sin\theta_2 \cos\theta_1 - \cos\theta_1\sin\theta_2  
\cos(\phi_1+\phi_2)) dr \,d\theta_1 \nonumber\\  
&&+ \frac 12 r(\sin\theta_1 \cos\theta_2 -  
\cos\theta_2\sin\theta_1  
\cos(\phi_1+\phi_2)) dr \,d\theta_2\nonumber\\  
&&+\frac12 r\sin\theta_1\sin\theta_2\sin(\phi_1+\phi_2) dr \,d(\phi_1+\phi_2)  
\nonumber\\  
&&+\fft{r^2}4\bigg((\cos\theta_1-\cos\theta_2) d\psi \,  
d(\phi_1+\phi_2)\nonumber\\  
&&\qquad+ \sin\theta_2\sin(\phi_1+\phi_2)d\psi \, d\theta_1  
-\sin\theta_1 \sin(\phi_1+\phi_2) d\psi\, d\theta_2\nonumber\\  
&&\qquad+ \sin\theta_1\cos\theta_2\sin(\phi_1+\phi_2)d\theta_2 \,  
d(\phi_1+\phi_2)\nonumber\\  
&&\qquad+ \sin\theta_2\cos\theta_1\sin(\phi_1+\phi_2)d\theta_1 \,  
d(\phi_1+\phi_2)\nonumber\\  
&&\qquad+(1-\cos\theta_1\cos\theta_2  
+\sin\theta_1\sin\theta_2 \cos(\phi_1+\phi_2))  
d(\phi_1+\phi_2){}^2\bigg)\bigg],\quad  
\eea  
where $w_4$ is given by (\ref{w4}).  
  
\subsection{A $\mathbb{P}^1$-inspired holomorphic embedding}  
  
Here we consider a different embedding which is inspired by the small  
resolution of the conifold \cite{candelas,resolution}. In this embedding the  
D7-branes are perpendicular to a $\mathbb{P}^1$ submanifold inside the  
conifold. The resolution of the conifold can naturally be described  
in terms of the $(z_1,z_2,z_3,z_4)$ of (\ref{zcoord}). In particular,  
resolving the conifold means replacing the equation  
$z_1z_2-z_3z_4=0$ by the pair of equations  
\begin{equation}  
\eqlabel{rconifold}  
\begin{pmatrix}  
z_1&z_3\\  
z_4&z_2  
\end{pmatrix}  
\begin{pmatrix}  
\l_1\\  
\l_2  
\end{pmatrix}=0,  
\end{equation}  
where the product $\l_1 \l_2 \not=0$. Note that $(\l_1,\l_2)\in  
\mathbb{CP}^1$ (any pair obtained from a given one  
by multiplication by a nonzero complex number is also a solution).  
Thus $(\l_1,\l_2)$ is uniquely characterized by the ratio $\l=\l_2/\l_1$  
in the region where $\l_1\ne 0$. Working on this patch, a solution to  
(\ref{rconifold}) takes the form%
\footnote{In the region where $\l_1$ is allowed to be zero we have  
instead $\l_2\ne 0$. In this case the general solution can be written as  
$\W=\begin{pmatrix}  
z_1&-z_1\m\\  
z_4&-z_4\m  
\end{pmatrix}$,  
where $\m=\l_1/\l_2$.}  
\begin{equation}  
\eqlabel{rw}  
\W=\begin{pmatrix}  
-z_3\l&z_3\\  
-z_2\l&z_2  
\end{pmatrix}.  
\end{equation}  
Thus $(z_2,z_3,\l)$ are the three complex coordinates characterizing  
the resolved conifold in the patch where $\l_1\ne 0$.  
  
As usual, the conifold metric is $g_{m{\bar n}}=\del_m\del_{\bar  
n}\calf$, where $\calf$ is the K\"ahler potential. In this  
case, the radial coordinate defined by (\ref{eq:rcoord}) takes  
the form  
\begin{equation}  
r^2=\mbox{tr} (\W^{\dagger} \W) =(1+|\l|^2)(|z_2|^2+|z_3|^2).  
\end{equation}  
The  K\"ahler potential for the conifold is simply%
\footnote{Note that in this specific patch the matrix (\ref{rw})  
satisfies the defining equation for the conifold. One can think of  
this matrix as a ``solution'' with $z_1=-\l z_3$ and $z_4=-\l z_2$.}  
\be  
\eqlabel{kpconrecoord}  
\calf=(r^2)^{2/3}=\bigg[(1+|\l|^2)(|z_2|^2+|z_3|^2)\bigg]^{2/3}.  
\ee  
In this parametrization it is natural to consider a modification of the  
K\"ahler potential that depends on the two variables:  
\be  
s=|z_2|^2+|z_3|^2, \qquad t=|\l|^2.  
\ee  
Doing so, we find that the determinant of the metric takes the form  
\be  
\eqlabel{tp1}  
\det g_{m\bar{n}}=\left(\calf_t\calf_s^2 + \calf_t\calf_s\calf_{ss} s -  
\calf_s\calf_{st}^2 t s + \calf_{tt}\calf_s^2 t +  
\calf_s\calf_{ss}\calf_{tt}t s\right).  
\ee  
As explained previously, our task is to  
find a solution $\calf(s,t)$ of equation  
(\ref{mastereqn}) with the condition that  
when the number of D7-branes goes to zero we recover the K\"ahler  
potential of the conifold.  
  
We now observe that the above equation is homogeneous in powers of $s$  
(as well as in powers of $t$).  Hence, although the equation is nonlinear,  
in principle all $s$ dependence could be removed by a judicious choice of  
$\calf\sim s^n$ for some power $n$.  
  
This suggests a separation of variables in $s$ and $t$ of the form  
$\calf = s^{n}f(t)$.  Eq.~(\ref{tp1}) becomes:  
\begin{equation}  
\mbox{det} g_{m\bar n}=n^3 s^{(3n-2)} f(f f'' t + f f' -(f')^2 t).  
\end{equation}  
Taking $n=2/3$ in the above gives an $s$ independent  
determinant, and leaves a non-linear ODE to solve:  
\begin{equation}  
\mbox{det}g_{m\bar n}=\left(\frac{2}{3}\right)^3 f(f f'' t+f f'-(f')^2 t)=  
\left(\frac{2}{3}\right)^{4}(1-n_f\log(t)).  
\eqlabel{eq:ftode}  
\end{equation}  
We have fixed $\Omega\bar{\Omega}=\left(2/3\right)^{4}$ by demanding that  
$f$ reduces to $f(t)=(1+t)^{2/3}$ for $n_f=0$ ({\it i.e.}~in the absence of  
D7-branes). Again, we have assumed that $\Omega$ does not change when $n_f$  
is nonzero. Although the source does not respect homogeneity in $t$, it  
is still useful to perform the change of variables  
\be  
t=\exp(y), \quad f(t)=\exp(y/3)l(y),  
\ee  
In this case, (\ref{eq:ftode}) implies that  
\be  
\eqlabel{ODE}  
l^3 \partial_y \partial_y \log(l)=(2/3)(1-n_f y).  
\ee  
The standard conifold case can be recovered  by taking  
$l=(2 \cosh(\fft{y}2))^{2/3}$. Interestingly, the combination  
$l=(2 A \cosh(\frac{y-y_0}{2 A}))^{2/3}$  
also gives a Ricci flat metric where the parameters are related to  
those discussed at length in \cite{pt}; however we take the simplest  
case with $A=1$, $y_0=0$.  
  
Note that, in fact, the $l^3$ prefactor in (\ref{ODE}) may be replaced  
by any non-zero power of $l$ through the replacement $l\to l^\alpha$  
where $\alpha$ is an arbitrary non-zero constant. Of particular interest  
is to take $l \rightarrow l^{2/3}$, because the conifold itself,  
(\ref{kpconrecoord}), has this $2/3$ power built in. This gives the equation  
\begin{equation}  
l^2\partial_y \partial_y \log(l)=(1-n_f y).  
\eqlabel{eq:newconi}  
\end{equation}  
The Ricci-flat solutions (corresponding to $n_f=0$) are now simply  
$l=2\cosh(\frac{y}{2})$.  Assuming a perturbative solution to  
(\ref{eq:newconi}) of the form  
\begin{equation}  
l=2 \cosh(\frac{y}{2})\exp(n_f g_1(y)+n_f^2g_2(y)+\cdots),  
\end{equation}  
we find that at each order, we must solve an equation of the form  
\begin{equation}  
L g_n(y) = S_n[g_1,\ldots,g_{n-1}],\qquad L = 1 + (1+\cosh y)\partial_y^2.  
\end{equation}  
Noting that $L$ admits the homogeneous solution  
\begin{equation}  
g(y) = c_1\tanh(\fft{y}2)+c_2\left(y\tanh(\fft{y}2)-2\right),  
\end{equation}  
we may construct the Green's function  
\begin{equation}  
G(y,y_0) = g_0(y)+\fft{(2-y_0\tanh\fft{y_0}2)\tanh\fft{y}2-(2-y\tanh\fft{y}2)  
\tanh\fft{y_0}2}{1+\cosh{y_0}}\theta(y-y_0),  
\end{equation}  
where $g_0(y)$ satisfies the homogeneous equation, and must be chosen to  
respect boundary conditions.  
  
At linear order in $n_f$, we find $Lg_1=-y/2$, which admits a simple  
solution  
\begin{equation}  
g_1=-\frac{y}{2}.  
\end{equation}  
Taking this form of $g_1$, the next order equation has source $Lg_2=-y^2/4$,  
and may be solved to yield  
\begin{equation}  
g_2=-2\log(1+e^y)+\fft{y(4-y)e^y}{2(1+e^y)}  
+\fft{1-e^y}{1+e^y}{\rm Li}_2(-e^y),  
\end{equation}  
up to the possible addition of a solution to the homogeneous equation.  
  
We have therefore managed, in this simple case, to find a  
modification to the K\"ahler potential of the conifold accommodating  
the presence of D7-branes up to second order (and in principle extendible  
to arbitrary order) in the number of D7-branes.  
Interestingly, however, as we will explain in the next section, this  
embedding of the D7-branes seems to be in  
a different class from the $w_4=0$ one.  
  
\section{Gauge theory}\label{gauge}  
  
Let us start by commenting on the decoupling limit.  
To make connection with gauge theories we first show that the  
supergravity solutions presented here allow for a region where only  
the low energy D-brane degrees of freedom are described. This is a necessary  
requirement to claim a duality between our supergravity backgrounds and  
the gauge theories appearing in the worldvolumes of branes.  
  
The Yang-Mills coupling of the gauge theory living on a D$p$-brane is  
\be  
g^2_{Dp}=g_s l_s^{p-3}.  
\ee  
We want to take the string scale to zero $l_s\to 0$ keeping the D3 SYM  
coupling fixed $g_{D3}^2=g_s$. This automatically means that  
$g^2_{D7}\to 0$ which implies that the D7-brane dynamics   
decouple, leaving us with a flavor symmetry.  
  
There are other quantities that we will keep fixed in the limit $l_s\to  
0$ such as the mass of the 7-7, 3-7 and 3-3 strings. This translates  
into a rescaling of the radial coordinate perpendicular to the D3-brane  
as $u={r}/{l_s^2}$.  This decoupling  
limit provides us with a consistency check on the solutions we  
find. In particular, we will require that when turning off the number  
of D7-branes the metric becomes either $AdS_5\times S^5$ for the D7 in  
flat space or $AdS_5\times T^{1,1}$ for D7's on the conifold.

The gauge theory of a stack of D7's on a D3 background has been  
discussed  in \cite{pg}. A complete account of the degrees of freedom  
of this system (and in general of D$p$/D$(p+4)$) is given in the textbook  
\cite{bigbook}.  
One interesting fact about the D3/D7 system is that the condition for  
the dilaton to be small, that is for the string loop corrections to  
be small requires  $r\ll e^{2\pi/(g_s N_f)}$. The condition to suppress  
the $\alpha'$ corrections, on the other hand, implies in the case of  
D3/D7 the radial distance from the D7 must satisfy $r\gg e^{-2\pi  
N_c/N_f}$. These two conditions have to be taken  
into consideration when interpreting the range of validity of the  
supergravity calculation. Generically, for large $g_s N_f$ we must  
consider only $N_c\gg N_f$. However, for very small $g_sN_f$, there is  
a range in which, in principle we could relax the relation between  
$N_c$ and $N_f$.

\subsection{Matching Pathologies: Naked singularities and Landau Poles}  
  
If one restricts to the form of the dilaton given in \ref{tauf} and  
rewrites the metric using $z_3=\rho e^{i\phi}$  
\be  
ds_6^2=dz_1d\bz_1+dz_2d\bz_2+  
\left(1-\frac{N_f}{4\pi }\log z_3\bz_3\right)dz_3d\bz_3,  
\ee  
it becomes clear that the metric has a naked singularity when   
\be  
e^{-\Phi}=1-\frac{N_f}{2\pi}\ln \left(\frac{\rho_3}{\rho_0}\right),   
\ee  
vanishes, that is, at:  
\be  
\rho_L=\rho_0\,\, e^{2\pi/ g_s N_f}.  
\ee  
A simple way to realize the relation to the Landau pole  
is through the identification  
\be  
\frac{1}{g_{YM}^2}=e^{-\Phi}.  
\ee  
Thus, the metric singularity coincides with the point at which  
$g_{YM}\to \infty$. It is, therefore, natural to consider only the  
region with $\rho_3 \le \rho_L$.   
  
\subsection*{Gauge theory dual of D3/D7/D7$'$}  
The gauge theory living on this  brane configuration has been described  
in the literature, for example in  \cite{afm,kk}. The analysis is closely related to a configuration of  
intersecting M5's discussed in \cite{koy}.  The  theory includes the $\caln=4$  
supermultiplet plus $N_f$ hypermultiplets $Q$ and $N_f'$ hypermultiplets  
$T$ transforming in the fundamental representation of SU$(N)$. A  
simple way to  
describe this theory is  via  
the superpotential  
\be  
W=W_{\caln=4}+\l_Q QZ_2\tilde{Q} + \l_T TZ_3\tilde{T},  
\ee  
where $Z_2$ and $Z_3$ are two of the three chiral superfields of the  
$\caln=4$ multiplet and we are explicitly using $\caln=1$ notation.  
  
The results of our supergravity analysis point to a generic appearance  
of nondiagonal terms in the metric. The explicit form of the metric  
(\ref{2stacksmixing}) shows a complicated pattern of mixing the  
$z_2$ and the $z_3$ coordinates. This seems to suggest that, at strong  
coupling, the two flavor sectors mix in a complicated way due to  
interactions.  
  
\subsection*{Gauge theory dual of D3/D7 on the conifold}  
  
D3-branes on the conifold \cite{kw} realize an $\caln=1$  
superconformal field theory (eight supercharges) with gauge group  
$\rm SU(N)\times SU(N)$. The matter content of this theory consists of  
chiral superfields $A_i$ and $B_i$ with $i=1,2$ transforming as $({\bf  
N},{\bf \bar{N}})$ and $({\bf \bar{N}},{\bf N})$ respectively. The  
fields $A$ and $B$ are doublets under one of the SU$(2)$'s in the  
global symmetry $\rm SU(2)\times SU(2)$. Under this  
symmetry a natural  
superpotential to add is \cite{kw}  
\be  
W_{conifold}=\frac{\l}{2} \epsilon^{ij}\epsilon^{kl}Tr A_iB_kA_jB_l.  
\ee  
  
As the example of $\caln=4$ shows, the structure of the deformation of  
the superpotential in the gauge theory is dictated by the geometry in  
the supergravity background.  In particular, note that the  
superpotential for the conifold is intimately related to the defining  
equation for the conifold (\ref{defining}) under the identification  
\be  
z_1=A_1B_1, \quad z_2= A_2B_2, \quad z_3 = A_1 B_2, \quad z_4=A_2 B_1.  
\ee  
To write down the superpotential  
corresponding to the embedding  
defined by $w_4=0$, recall that $w_4 \sim z_1+z_2$ and therefore  
the natural deformation of the above superpotential  
is of the form  
\be  
\eqlabel{spcon}  
W=W_{conifold}+ \lambda_Q  Q(A_1B_1+A_2B_2)\tilde{Q} +\lambda_{\tilde q}  
q(B_1A_1+B_2A_2)\tilde q.  
\ee  
In \cite{peter} a deformation of this type was motivated using the  
theory of D3-branes on $\mathbb{C}^2/\mathbb{Z}_2$.

\subsection*{Comments on the $\beta$-functions}  
The exact $\beta$ function for the canonical  
gauge coupling of $\caln=1$ with gauge group $SU(N)$ and $F$ flavors  
is \cite{beta}  
\be  
\eqlabel{nsvz}  
\beta(g)=-\frac{g^3}{16\pi^2}  
\frac{3N-F(1-\g)}{1-N\frac{g^2}{8\pi^2}},  
\ee  
where $\g$ is the anomalous dimension of the matter fields. One of the  
most attractive features of the solutions describing branes  
on the apex of the conifold is the precise geometrical description of the  
NSVZ $\beta$-function \cite{kw,kn}.  
  
One very important feature of (\ref{nsvz}) is that it serves as a guide  
in determining the boundary conditions of the supergravity solution  
describing flavor. Namely, as describe by equation (\ref{odilaton})  
the dilaton can be any holomorphic function of the coordinates. In  
particular the sign of the coefficient of the term containing the  
logarithm is arbitrary in the supergravity solution. In this case  
equation (\ref{nsvz}) dictates the sign.  
  
Another important remark about (\ref{nsvz}) is that it also allows to  
understand the region of the radial coordinates where the theory is  
describing the flavor sector. Note that in general the dilaton goes  
from a constant at infinity to a logarithmic term near the D7-branes.  
As we will see in what follows for the case of the conifold,  
only the logarithmic region (decompactification limit) contributes to  
the $\beta$-function in the expected way.  
  
For the Klebanov-Witten theory the assignment of  anomalous dimension of  
$\g= -1/2$ to the operators  
${\rm Tr} A_i B_j$ guarantees that the  beta function is zero, which  
corresponds to a superconformal fixed point \cite{kw}  
  
\be  
\frac{d}{d\log (\L/\m)}\frac{8\pi^2}{g_1^2}\approx  3N-2N(1-\g).  
\ee  
The two gauge couplings are related to the supergravity background as  
\cite{kw,ks}:  
\bea  
\frac{4\pi^2}{g_1^2}+\frac{4\pi^2}{g_1^2}&=&\frac{\pi}{g_s}e^{-\Phi},  
\nonumber \\  
\frac{4\pi^2}{g_1^2}-\frac{4\pi^2}{g_1^2}&=&\frac{1}{2\pi  
\a'}\left(\int\limits_{S^2}B_2\right) -\pi \quad {\rm mod}\,\, 2\pi.  
\eea  
The case with probe D7-branes was considered in \cite{peter}.  
In the  absence of $B_2$ field we obtain that the two couplings are  
the same. To read off their value we need the explicit expression for  
$w_4$ which can be obtained from (\ref{defining}),  
(\ref{wparametrization}) and (\ref{su2})  
\be  
w_4=\frac{i\,r}{\sqrt{2}}e^{\frac{i}{2}(\psi+\p_1+\p_2)}  
\big[\cos\frac{\te_1}{2}\sin\frac{\te_2}{2}-  
\sin\frac{\te_1}{2}\cos\frac{\te_2}{2} e^{-i(\phi_1+\phi_2)}\big]\eqlabel{w4}.  
\ee  
Assume, that the renormalization energy scale is proportional  to the radius  
$\Lambda \sim \r$ (the perpendicular distance from the D3-branes)  
we have the following expression for the running of the coupling  
\be  
\frac{\del}{\del  
\Lambda}\frac{8\pi^2}{g_{YM}^2}=-\frac{3}{2}N_f(1-\g_Q),  
\ee  
The value of $\g_Q$ can be found, similarly to a computation of  
\cite{peter}, by power counting. Given that the dimension of the  
superpotential (\ref{spcon}) is three, and that the dimension of the fields  
$A$'s and $B$'s is 3/4 as argued in \cite{kw}, we conclude that  
$\g_Q=1/4$.  
  
For the case of the $\mathbb{P}^1$ parametrization of the conifold we  
note that  $\tau\sim \log \lambda$ and  
\be  
\lambda=\tan\frac{\te_2}{2}e^{-i\phi_2}.  
\ee  
This combination is independent of the radial coordinate $\rho$,  
therefore,  
keeping the natural identification of $\r$ with the RG energy scale $\Lambda$  
we conclude that this deformation of the gauge theory does not affect  
the $\beta$-function and therefore does not truly amount to adding  
flavor to the KW gauge theory.  
  
\section{Conclusions}\label{conclusions}  
  
We conclude with a list of open problems which we believe would be  
interesting to address. The first, and perhaps most haunting problem in the  
gauge/gravity correspondence is how to achieve asymptotic freedom, or in  
other words how to follow the theory from strong coupling where  
supergravity is a good approximation into the weak coupling regime.  
This problem is well beyond the scope of the present paper. But  
it must be kept in mind in order to understand the limitations of  
the present work. Some direct open problems resulting from  
our investigation are:  
  
$\bullet$ Application of our techniques to other K\"ahler spaces. Most  
prominently, the resolved and deformed conifolds.  
  
$\bullet$ The inclusion of 3-form fluxes. In the case of the conifold  
we refer to $G_3$ which corresponds to including fractional branes  
which change the relative rank of the two gauge groups.  
  
$\bullet$ An interesting open problem would be to understand the  
backreaction for theories that do not necessarily admit a formulation  
exploiting K\"ahler geometry. Such seems to be the case for the  
Maldacena-N\'u\~nez solution and the D4/D6 bound state.  
  
$\bullet$  
Ideally, it would be desirable to obtain the spectrum of the  
different modes for the class of solutions we have presented. We  
should point out that some analysis of the spectrum (for example of  
mesons) has been presented in the probe approximation  
\cite{cobi,martin1,martin2,carlos}. But an analysis of the fully backreacted  
case will certainly yield a more robust picture.  In this paper we took  
a step further by showing how the backreacted D3/D7 solution could be used  
to derive the scalar (meson) spectrum of a ${\cal N}=1$ gauge theory  
realized on the worldvolume of a D7$'$ probe.  
  
To conclude, we believe we have explicitly demonstrated a viable  
way to construct fully backreacted supergravity backgrounds dual to  
four-dimensional $\caln=1$  SYM with flavor in the fundamental  
representation. We hope to return to many of the fascinating issues in  
the gauge/gravity correspondence with flavor in future work.  
  
\section*{Acknowledgments}  
We would like to thank L. Anguelova, S. Monni and M. Stephanov for comments  
and C. N\'u\~nez for discussions and collaboration  
in a related project. We would also like to acknowledge useful  
discussions  with participants of the Trento workshop, most notably:   
A. Armoni, J. Barb\'on, A. Buchel and  D. Mateos. B.A.B. would like to thank R. McNees for many  
useful conversations. LAPZ would like to thank A. Hashimoto and  
M. Mahato for many clarifications in the course of a collaboration in  
related topics.   
This work was supported in part by the US Department of Energy  
under grant DE-FG02-95ER40899. D.V. was supported by DOE grant  
DE-FG02-91ER40671.  
  
\appendix  
\section{Comments on the Monge-Amp\`ere equation}\label{mam}  
In this appendix we collect some known results about the  
Monge-Amp\`ere equation that are relevant for the analysis in the main  
text.  The literature on this subject is vast \cite{ma,ma2,ma3}  
and we refer the reader to those references for further details.  
  
We start with some motivation of why Monge-Amp\`ere equations are  
relevant for our analysis.  We assume (four-dimensional) Poincar\'e  
symmetry and (for simplicity) turn off 3-form flux.  
As in section \ref{d3d7}, one arrives at equation (\ref{mastereqn})  
for type IIB supergravity  
\begin{equation}  
\eqlabel{MACP}  
\det(\partial_{\mu}\partial_{\bar{\nu}} \calf)=\Omega \bar{\Omega}\,  
\Im(\tau).  
\end{equation}  
To place this in the proper context of the Monge-Amp\`ere equation, we  
must change the differentiation appearing above into differentiation  
with respect to real variables.  First, we consider the  
``decompactification'' limit of seven branes \cite{gsvy}, where  
$\tau \sim i C + i a \log(z)$.  We will then place our D3's close  
enough to the D7's so that this approximation is valid. Thus  
$\Im{(\tau)} \sim \log(z\bar{z})$ where we have chosen our origin so  
that $b=0$.  We then make the Ansatz that the functional dependence of  
the K\"ahler potential is $\calf \left(\log(z_1 \bar{z_1}),\ldots,  
\log(z_n \bar{z_n})\right)$, where the $\log$'s have been inserted  
for later convenience.  This requires that $\Omega$ has the form  
$\Omega=\Pi z_i^{m_i}$ for any set of $m$'s so that both left and right  
hand side are functions of the new variables $y_n=\log(z_n\bar{z}_n)$.  
Under this change Eq.~(\ref{MACP}) becomes:  
\begin{eqnarray}  
\label{hessian}  
\frac{\det(\partial_{i}\partial_{j} \calf)}{\Pi \exp{(y_i)}} &=& \Omega  
  \bar{\Omega}\,\Im(\tau),  
\nonumber\\  
\det(\partial_{i}\partial_{j} \calf)&=&\exp{\left(\Sigma  
(m_i+1)y_i\right)}\Im(\tau).  
\end{eqnarray}  
Note that the change of variables has introduced only a product of the  
$z$'s to powers.  This  
can be absorbed into the definition of $\Omega$, and so has no affect on  
the curvature. Equation (\ref{hessian}) is the standard Monge-Amp\`ere  
equation in real variables.  We now turn to some of the main results from  
the study of Monge-Amp\`ere equations.  
  
The Dirichlet problem for Monge-Amp\`ere equations is formulated as  
follows. Let $\Omega$ be a bounded strictly convex domain in  
$\reals^n$ $(n\ge 2)$ defined by a $C^\infty$ strictly convex function  
$h$  on $\bar\Omega$ satisfying $h|_{\del\Omega}=0$. Given $u(x)$, a  
$C^\infty$ function on $\del\Omega $ which is the restriction to  
$\del\Omega $  of a $C^\infty$ function $\gamma$ on $\bar\Omega$, we  
consider the equation:  
\be  
\eqlabel{mae}  
\log \det (\partial_{i}\partial_{j} \phi)= f(x, \phi), \qquad  
\phi|_{\del\Omega}=u,  
\ee  
where $f(x,t)\in C^\infty (\bar\Omega,\reals)$. The main result that  
we will use is an existence theorem which can be  stated as follows  
\cite{ma,ma2,ma3}. The Dirichlet problem  
(\ref{mae}) has a unique solution belonging to $C^\infty(\bar\Omega)$,  
when $n=2$, if there exists a strictly convex upper solution $\g_0\in  
C^2(\bar\Omega)$ satisfying  
\be  
\det(\del^2_{ij} \g_0)\le \exp f(x,\g_0), \qquad \g_0|_{\del\Omega}=u,  
\ee  
and if $f'_t(x,t)\ge 0$  for all $x\in \Omega$ and $t\le  
\sup_{\del\Omega}u$.

  
\end{document}